\documentclass[aps,prx,twocolumn,superscriptaddress]{revtex4-2}
\usepackage{graphicx}
\usepackage{dcolumn}
\usepackage{bm}
\usepackage{color}
\usepackage{ulem}
\usepackage[usenames,dvipsnames,svgnames,table]{xcolor}

\usepackage{amsfonts}       
\usepackage{amsmath}        
\usepackage{bm}             
\usepackage{mathrsfs}       
\usepackage{setspace}       
\usepackage{microtype}     
\usepackage{enumerate}
\usepackage{soul}      
\usepackage{xspace}
\usepackage[unicode=true,
  linktocpage,
  linkbordercolor={0.5 0.5 1},
  citebordercolor={0.5 1 0.5},
  linkcolor=blue]{hyperref}
\usepackage{changes}
\usepackage{makecell}

\newcommand {\vecck}{{\vec k}}  
\newcommand {\veck}{{\vec k}}  

\newcommand{\pstar}{$p^{\star}$\xspace}

\begin{document}

\title{Seebeck coefficient in a cuprate superconductor: particle-hole asymmetry in the
strange metal phase and Fermi surface transformation in the pseudogap phase}

\author{A. Gourgout}
\thanks{These two authors contributed equally to this work}
\affiliation{Institut Quantique, D\'epartement de physique \& RQMP, Universit\'e de Sherbrooke, Sherbrooke, Qu\'ebec, Canada}

\author{G. Grissonnanche}
\thanks{These two authors contributed equally to this work}
\affiliation{Institut Quantique, D\'epartement de physique \& RQMP, Universit\'e de Sherbrooke, Sherbrooke, Qu\'ebec, Canada}
\affiliation{Laboratory of Atomic and Solid State Physics, Cornell University, Ithaca, NY, USA}
\affiliation{Kavli Institute at Cornell for Nanoscale Science, Ithaca, NY, USA}
\email{gael.grissonnanche@cornell.edu}

\author{F. Lalibert\'e}
\affiliation{Institut Quantique, D\'epartement de physique \& RQMP, Universit\'e de Sherbrooke, Sherbrooke, Qu\'ebec, Canada}

\author{A. Ataei}
\affiliation{Institut Quantique, D\'epartement de physique \& RQMP, Universit\'e de Sherbrooke, Sherbrooke, Qu\'ebec, Canada}

\author{L. Chen}
\affiliation{Institut Quantique, D\'epartement de physique \& RQMP, Universit\'e de Sherbrooke, Sherbrooke, Qu\'ebec, Canada}

\author{S. Verret}
\affiliation{Institut Quantique, D\'epartement de physique \& RQMP, Universit\'e de Sherbrooke, Sherbrooke, Qu\'ebec, Canada}

\author{J.-S. Zhou}
\affiliation{Materials Science and Engineering Program/Mechanical Engineering, University of Texas - Austin, Austin, TX, USA}

\author{J. Mravlje}
\affiliation{Department of Theoretical Physics, Institute Jo\v{z}ef Stefan, Ljubljana, Slovenia}

\author{A. Georges}
\affiliation{Coll{\`e}ge de France, 11 place Marcelin Berthelot, Paris, France}
\affiliation{Center for Computational Quantum Physics, Flatiron Institute, New York, NY, USA}
\affiliation{CPHT, CNRS, Ecole Polytechnique, IP Paris, Palaiseau, France}
\affiliation{DQMP, Universit{\'e} de Gen{\`e}ve, Geneva, Switzerland}

\author{N. Doiron-Leyraud}
\affiliation{Institut Quantique, D\'epartement de physique \& RQMP, Universit\'e de Sherbrooke, Sherbrooke, Qu\'ebec, Canada}

\author{Louis~Taillefer}
\email[]{louis.taillefer@usherbrooke.ca}
\affiliation{Institut Quantique, D\'epartement de physique \& RQMP, Universit\'e de Sherbrooke, Sherbrooke, Qu\'ebec, Canada}
\affiliation{Canadian Institute for Advanced Research, Toronto, Ontario, Canada}

\date{\today}

\begin{abstract}

We report measurements of the Seebeck effect in both the $ab$ plane ($S_{\rm a}$) and along the $c$ axis ($S_{\rm c}$) of the cuprate superconductor La$_{1.6-x}$Nd$_{0.4}$Sr$_{x}$CuO$_4$ (Nd-LSCO), performed in magnetic fields large enough to suppress superconductivity down to low temperature. We use the Seebeck coefficient as a probe of the particle-hole asymmetry of the electronic structure across the pseudogap critical doping $p^{\star} = 0.23$.
Outside the pseudogap phase, at $p = 0.24 > p^{\star}$, we observe a positive and essentially isotropic Seebeck coefficient as $T \rightarrow 0$. That $S > 0$ at $p = 0.24$ is at odds with expectations given the electronic band structure of Nd-LSCO above $p^{\star}$ and its known electron-like Fermi surface.
We can reconcile this observation by invoking an energy-dependent scattering rate with a particle-hole asymmetry,
possibly rooted in the non-Fermi liquid nature of cuprates just above $p^{\star}$.
Inside the pseudogap phase, for $ p < p^{\star}$, $S_{\rm a}$ is seen to rise at low temperature as previously reported, consistent with the drop in carrier density $n$ from $n \simeq 1 + p$ to $n \simeq p$ across $p^{\star}$ as inferred from other transport properties.
In stark contrast, $S_{\rm c}$ at low temperature becomes negative below $p^{\star}$, a novel signature of the pseudogap phase.
The sudden drop in $S_{\rm c}$ reveals a change in the electronic structure of Nd-LSCO upon crossing $p^{\star}$. We can exclude a profound change of the scattering across $p^{\star}$ and conclude that the change in the out-of-plane Seebeck coefficient originates from a transformation of the Fermi surface.

\end{abstract}

\maketitle

\section{Introduction}

The pseudogap phase of cuprate superconductors, in particular its link with the high temperature superconductivity, remains an enduring mystery of condensed matter physics.
While no clear phase transition at its characteristic temperature $T^{\star}$ is observed in transport and thermodynamic properties, the low temperature crossing of its critical doping $p^{\star}$ in the absence of superconductivity yields clear signatures of a transition~\cite{proust2019}.
The normal-state Hall coefficient of YBa$_2$Cu$_3$O$_y$ (YBCO)~\cite{badoux2016}, La$_{1.6-x}$Nd$_{0.4}$Sr$_{x}$CuO$_4$ (Nd-LSCO)~\cite{collignon2017} and Bi$_2$Sr$_{2-x}$La$_{x}$CuO$_{\rm 6+\delta}$ (Bi2201)~\cite{lizaire2020transport} shows that the carrier density $n$ changes abruptly from $n \simeq 1 + p$ to $n \simeq p$ when crossing $p^{\star}$ from above.
This is also observed in the electrical resistivity of La$_{2-x}$Sr$_x$CuO$_4$ (LSCO)~\cite{laliberte2016} and Nd-LSCO~\cite{collignon2017}, and the thermal conductivity (at $T\rightarrow 0$) of Nd-LSCO~\cite{michon2018}.
Clear evidence for a transformation of the Fermi surface of Nd-LSCO across $p^{\star}$ was recently obtained from angle-dependent magneto-resistance (ADMR)~\cite{fang2020}. The low temperature electronic specific heat \cite{michon2019,girod2021normal} shows a logarithmic divergence at $p^{\star}$, evidence for a quantum phase transition.

Recently, the thermopower was used as a complementary probe of the carrier density at low temperatures across $p^{\star}$ in Nd-LSCO~\cite{collignon2021thermopower}.
Indeed, in the $T = 0$ limit and for a single parabolic band, the Seebeck coefficient depends on only two parameters: the coefficient of electronic specific heat $\gamma = C_{\rm el}/T$ and the carrier density $n$ ($e$ is the electron charge):
\begin{equation}
\frac{S}{T} = \frac{\gamma}{ne}.
\label{eq:behnia_eq}
\end{equation}
According to this expression, the ratio $S/T$ can therefore be seen as the specific heat per carrier.
Simplified as it is, this physical picture was shown by Behnia \textit{et al.}~\cite{behnia2004}
to account empirically for the observed low-temperature value of the Seebeck coefficient for a great variety of materials that includes common metals and several strongly correlated materials.

In Nd-LSCO, Eq.~(\ref{eq:behnia_eq}) articulates that the sudden increase in the in-plane Seebeck
coefficient $S_{\rm a}/T$ as $p$ crosses below $p^{\star}$~\cite{collignon2021thermopower} is consistent with the drop in $n$ at $p^{\star}$ inferred from Hall effect, electrical resistivity, and thermal conductivity measurements. Although the relative change in $S_{\rm a}/T$ across $p^{\star}$ seems well captured by Eq. (\ref{eq:behnia_eq}), it is unable to explain the positive sign of the in-plane Seebeck coefficient, which is at odds with the electron-like band of Nd-LSCO above $p^{\star}$~\cite{verret2017}. To date, the sign of the Seebeck coefficient in overdoped cuprates remains a mystery.

At a more fundamental level, the Seebeck coefficient is actually a ratio of transport coefficients which
involves states immediately below and above the Fermi level, in contrast with electrical transport
that is only sensitive to the properties of the Fermi surface.
As a result, the Seebeck coefficient is controlled by the particle-hole asymmetry between
occupied and unoccupied states around the Fermi level \cite{behnia_2015, kondo2005}.
This asymmetry can originate both from the dispersion of electronic excitations (band structure)
and from the energy dependence of the scattering rate.
Accounting for these effects clearly goes beyond the simplified expression
Eq.~(\ref{eq:behnia_eq}).

In the present article, we present measurements of the out-of-plane Seebeck coefficient $S_{\rm c}$. We report a sudden qualitative change of $S_{\rm c}$ in Nd-LSCO across $p^{\star}$. We show that in order to successfully describe the behaviour of the Seebeck coefficient along both directions, it is crucial to  take into account the  particle-hole asymmetry (`skewness') in the energy dependence of the scattering rate.

\begin{figure}[t!]
\includegraphics[width=0.45\textwidth]{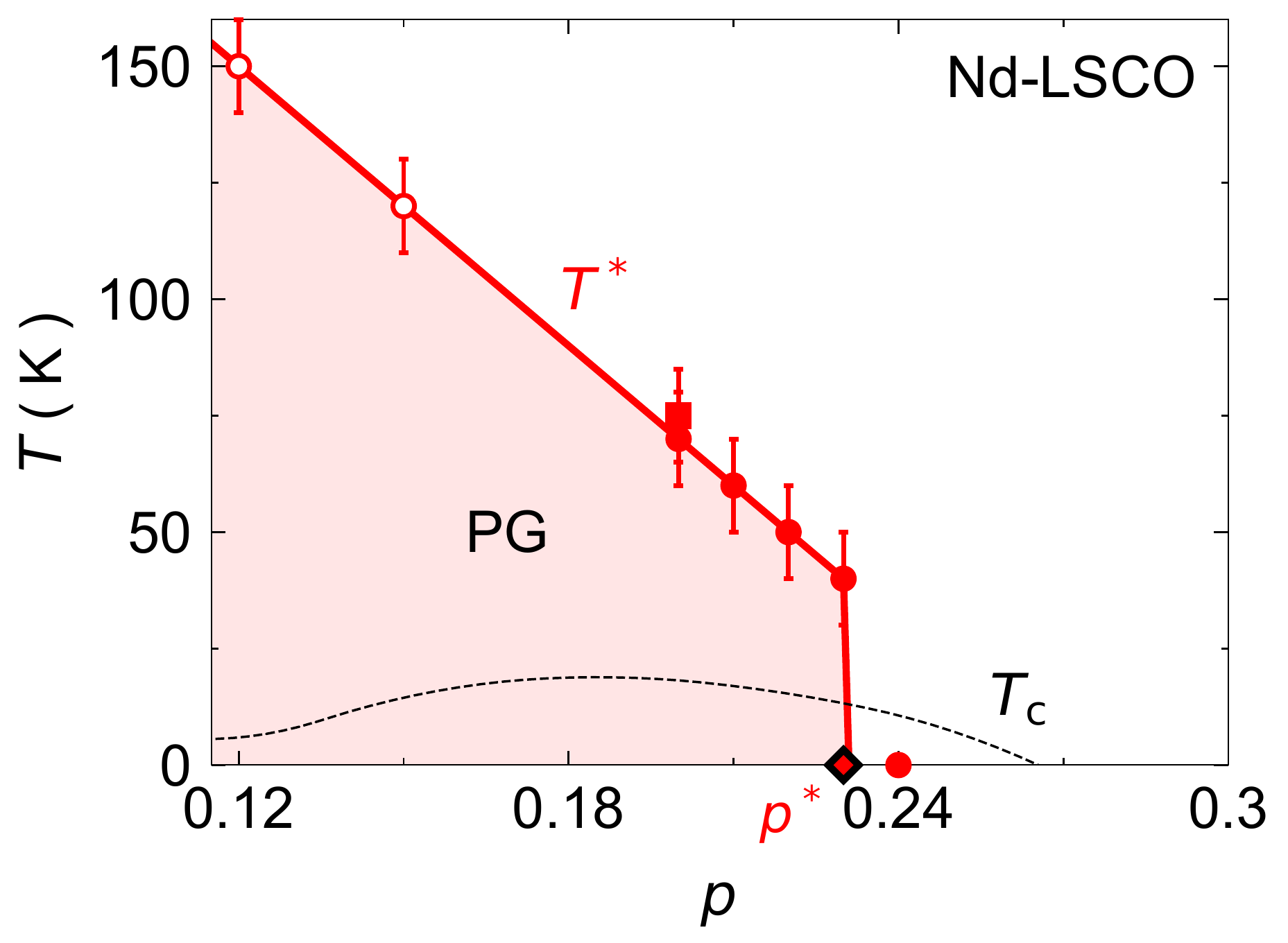}
\caption{Temperature-doping phase diagram of the cuprate superconductor Nd-LSCO. Red points represent the temperature below which the pseudogap phase appears, $T$$^{\star}$, obtained from resistivity \cite{daou2009,collignon2017} (full points show the dopings studied here) and ARPES~\cite{matt2015} (full square) measurements. The red shaded area represents the pseudogap (PG) phase, which disappears at the critical doping $p^{\star} = 0.23 \pm 0.01$ (red diamond). The full line is a guide to the eye. The superconducting phase (in zero field) is bounded by $T_{\rm c}$ (dashed line).
}
\label{fig:diagram}
\end{figure}

Nd-LSCO is a single-layer, tetragonal cuprate superconductor with low critical temperature $T_c$ and field $H_{c2}$, making it an ideal candidate to study the field-induced normal state down to low temperatures.
Its phase diagram is shown in Fig.~\ref{fig:diagram}, where the pseudogap temperature $T^{\star}$ extracted from resistivity measurements~\cite{collignon2021thermopower} is displayed and seen to be in agreement with angle-resolved photoemission spectroscopy (ARPES) measurements~\cite{matt2015}.

Above $p^{\star}$, at $p = 0.24$, our measurements reveal that $S_{\rm a}$ and $S_{\rm c}$ are positive and essentially equal at low temperature, both ending with a $\log(1/T)$ dependence below about 10~K. Although it is consistent with Eq. (\ref{eq:behnia_eq}), the isotropy of the Seebeck coefficient at $p=0.24$ as $T\rightarrow 0$ cannot be understood in terms of the band structure since it predicts the wrong magnitude of $S_{\rm c}$, and is of the wrong sign for $S_{\rm a}$. We show that taking into account the angle dependence and temperature dependence of the scattering rate deduced from ADMR~\cite{grissonnanche2020b} and adding a linear energy dependence that is particle-hole asymmetric reconciles the calculated and measured Seebeck coefficients along both directions.

Upon crossing into the pseudogap phase at $p^{\star}$, we find that $S_{\rm c}$ becomes negative at low temperature, in contrast to $S_{\rm a}$ which remains positive. This contrasts with Eq. (\ref{eq:behnia_eq}) that predicts that the drop in $n$ at $p^{\star}$ should also be observed in the interlayer $c$ axis component $S_{\rm c}$.
The negative $S_{\rm c}$ reflects a profound transformation of the electronic structure across $p^{\star}$.

\section{Methods}

Single crystals of Nd-LSCO were grown by a traveling solvent/floating zone technique in an image furnace, with a Nd content of 0.4 and doping $p$ = 0.20, 0.21, 0.22, 0.23 and 0.24 (for more details see Collignon et al. \cite{collignon2017}).
These were subsequently cut into bar-shaped samples with typical dimensions of 1 mm x 0.5 mm x 0.2 mm.
For each doping, separate samples with their length respectively along the $a$ axis and the $c$ axis were cut.

We measured the Seebeck coefficient using an AC-technique (Alternative Current) that was originally developed for thin films~\cite{wang2019} and that we have adapted for measuring bulk materials.
An AC thermal excitation was generated through the sample by sending an electric current at a frequency $\omega \sim 1$~Hz to a 5k$\Omega$ strain gauge used as a heater. This generates a longitudinal thermal gradient $\Delta T_{\rm AC}$ along the length of the sample, measured at a frequency $2\omega$ using two absolute type E thermocouples.
In response to that thermal gradient, a Seebeck voltage $\Delta V_{\rm AC}$ is measured at a frequency $2\omega$ with phosphor-bronze wires using the same contacts as for $\Delta T_{\rm AC}$, which eliminates uncertainties associated with the geometric factor.
The Seebeck coefficient is then given by $S = - \Delta V_{\rm AC} / \Delta T_{\rm AC}$.
Our method differs from previous AC Seebeck measurements by accounting for both the modulus and the phase of the thermal and thermoelectric signals.
This technique can be in principle extended to all orders in frequencies.
The advantages of using an AC method over DC is that the DC technique requires to measure the Seebeck coefficient at constant temperature steps and wait the necessary time to reach equilibrium, which can take minutes and even hours closer to room temperature, it is also limited in resolution to few milliKelvins fluctuations for the thermal gradient and few nanoVolts for the Seebeck voltage.
The AC Seebeck technique on the other hand does not require for the sample to reach equilibrium, which makes the measurements quasi-instantaneous, and comes down to a resolution in the $\sim 10$~$\mu$K range because of the use of lock-in amplifiers.
However, it takes a real time measurement of the sample temperature, which comes with its own set of challenges.

The thermocouple and Seebeck voltages were amplified using EM Electronics A10 preamplifiers and picked-up using SR830 lock-in amplifiers at the thermal excitation frequency $2\omega$.
Our method allows us to measure $S(T)$ continuously from 2 to 300K within a few hours and with a much better signal to noise ratio compared to a standard steady-state DC technique. It was carefully benchmarked against a steady-state DC method on several samples.\\

\begin{figure}[t!]
\includegraphics[width=0.45\textwidth]{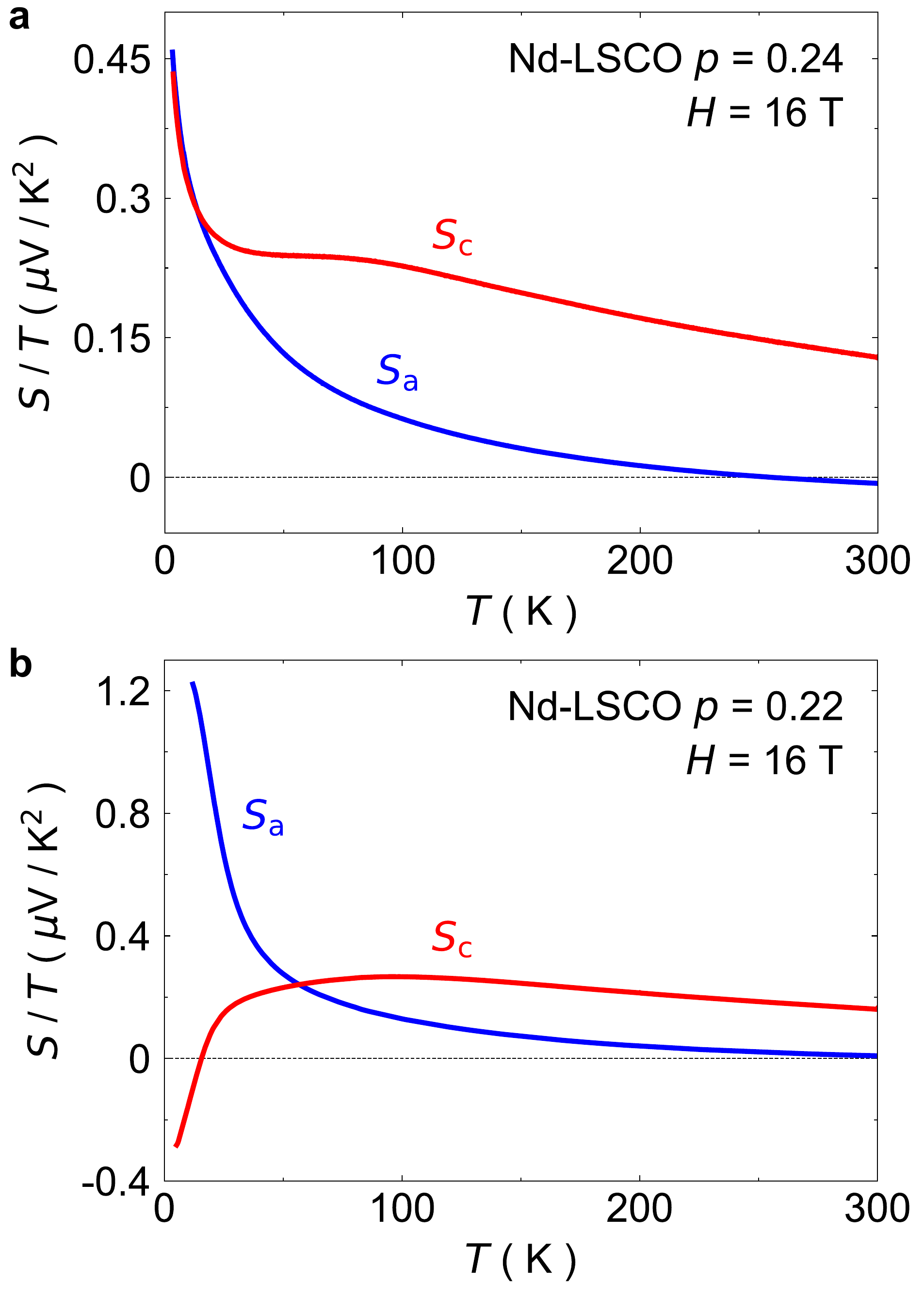}
\caption{In-plane ($S_{\rm a}$, blue) and out-of-plane ($S_{\rm c}$, red) Seebeck coefficient of Nd-LSCO, plotted as $S/T$ vs T at $H=16$~T for (\textbf{a}) $p = 0.24>p^{\star}$ and (\textbf{b}) $p~=~0.22~<~p^{\star}$.}
\label{fig:SoT_0p24_0p22}
\end{figure}

\section{Results}

\subsection{Outside the pseudogap phase $p > p^{\star}$}

In Fig.~\ref{fig:SoT_0p24_0p22}(a) we show $S_{\rm a}/T$ and $S_{\rm c}/T$ as a function of temperature in Nd-LSCO just outside the pseudogap phase, at $p = 0.24$, in the normal state induced by an applied field of $H = 16$~T ($H_{\rm c2} = 9$~T in Nd-LSCO and Eu-LSCO at $p = 0.24$ \cite{michon2019}).
We observe that $S_{\rm a}$ is negative at room temperature, similar to what is observed in overdoped Bi2201~\cite{kondo2005}, for example.
Upon cooling, $S_{\rm a}/T$ increases and changes sign around 250 K, reaching a positive value of about 0.45 $\mu$V/K$^{2}$ at the lowest measured temperature $T \simeq 2$~K.
Conversely, $S_{\rm c}/T$ starts from a positive value at 300~K, it increases until it reaches a plateau around 60 K, but at lower temperature starts to increase again.
Below about 10~K a remarkable isotropy appears: $S_{\rm a}/T$ and $S_{\rm c}/T$ are essentially identical despite the quasi two-dimensional character of the
electronic structure and associated anisotropy in the electrical resistivity (in Nd-LSCO at $p = 0.24$ the $c$ axis resistivity $\rho_c$ is approximately 250 times larger than $a$ axis resistivity $\rho_a$~\cite{daou2009}).

\begin{figure}[t!]
\includegraphics[width=0.45\textwidth]{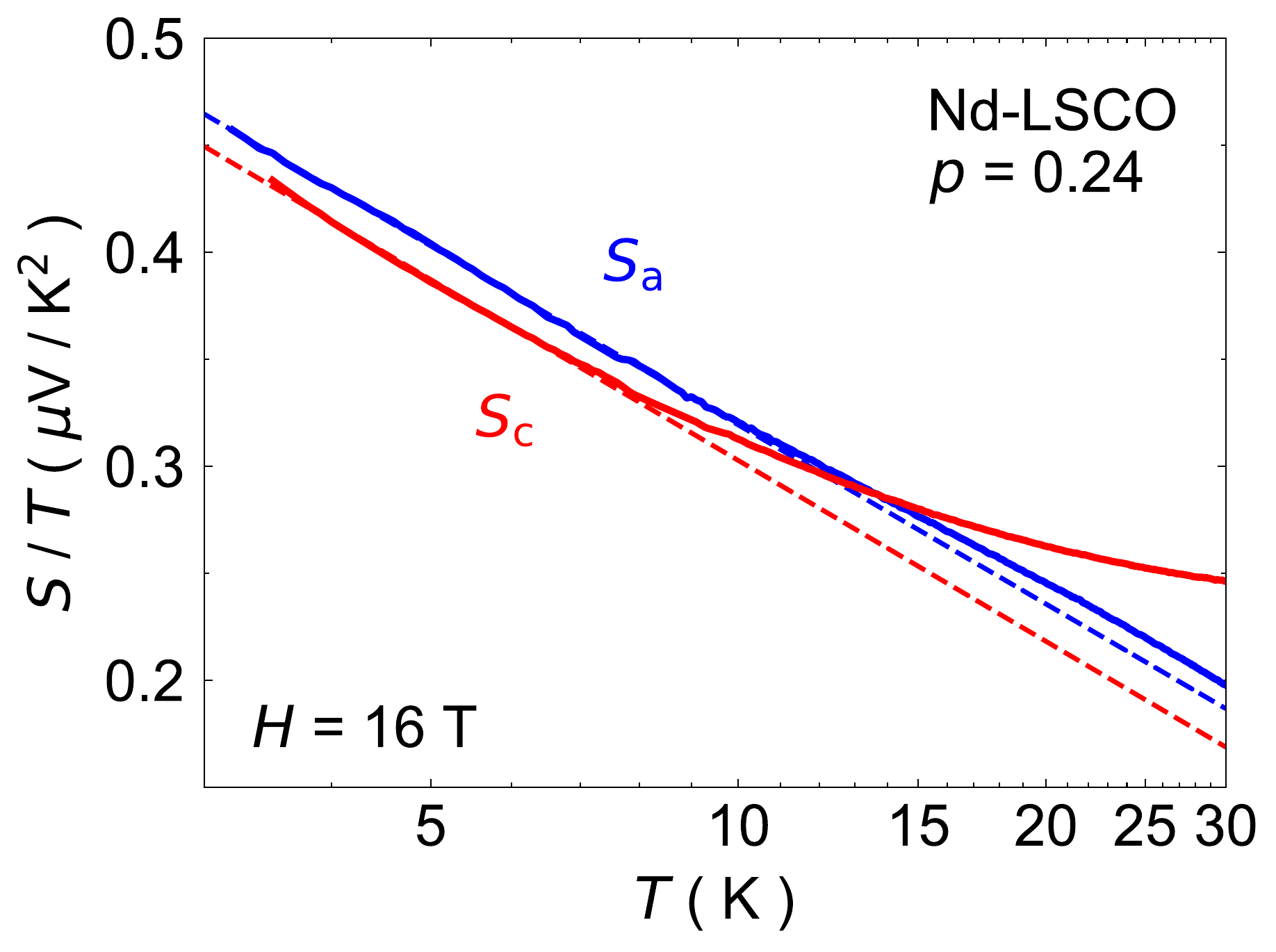}
\caption{In-plane ($S_{\rm a}$, blue) and out-of-plane ($S_{\rm c}$, red) Seebeck coefficients of Nd-LSCO at $p = 0.24> p^{\star}$ plotted as $S/T$ vs $\log(T)$ at $H=16$~T. Note the logarithmic dependence below $\sim 10$~K, where $S$ is essentially isotropic ($S_{\rm a} \simeq S_{\rm c}$).}
\label{fig:logT_0p24}
\end{figure}

As seen in Fig. \ref{fig:logT_0p24}, zooming in on the low temperature data reveals that $S_{\rm a}/T$ and $S_{\rm c}/T$ both display a $\log(1/T)$ dependence below $\sim 10$~K, as previously reported for $S_{\rm a}/T$ in Nd-LSCO~\cite{daou2009S}, Eu-LSCO~\cite{laliberte2011} at $p~=~0.24$ and in Bi2201 near $p^{\star}$~\cite{lizaire2020transport}.
This behavior is also observed in the electronic specific heat of Nd-LSCO and Eu-LSCO at $p = 0.24$ whereby $C_{\rm el} / T \sim \log(1/T)$ below $\sim 10$~K is interpreted as a signature of quantum criticality in the vicinity of the pseudogap critical point~\cite{michon2019,girod2021normal}.
Another typical signature of quantum criticality is a $T$-linear dependence of the electrical resistivity as $T \to 0$ observed in Nd-LSCO at $p = 0.24$ both in the plane and along the $c$ axis~\cite{daou2009}.

\begin{figure}[t]
\includegraphics[width=0.42\textwidth]{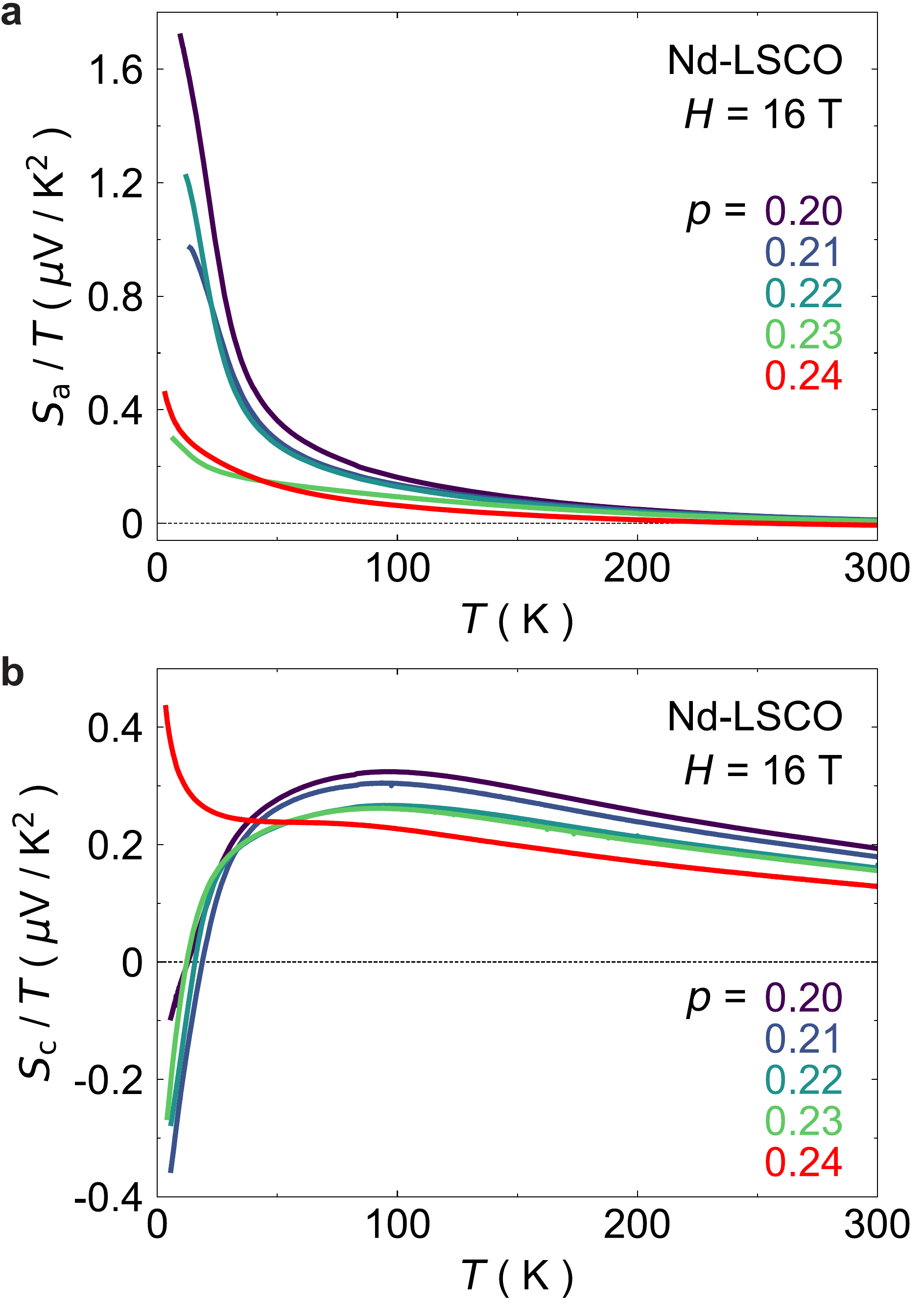}
\caption{(\textbf{a}) $S_{\rm a}/T$ and (\textbf{b}) $S_{\rm c}/T$ as a function of temperature, at $H=16$~T, in Nd-LSCO at dopings as indicated, on both sides of the pseudogap critical point $p^\star = 0.23$. Note the qualitative change in $S_{\rm c}$ upon crossing below $p^\star$, whereby $S_{\rm c}$ suddenly becomes negative at low $T$.}
\label{fig:SoT_doping}
\end{figure}

\subsection{Inside the pseudogap phase $p < p^{\star}$}

Now going inside the pseudogap phase, we see in Fig.~\ref{fig:SoT_doping}(a) that $S_{\rm a}/T$ undergoes a large enhancement at low temperature for $p < 0.23$.
This was previously reported and discussed in detail in ref.~\cite{collignon2021thermopower} and here we show the continuous $T$-dependent curves of $S_{\rm a}/T$ in $H = 16$~T obtained using our AC technique (our AC data are in excellent agreement with the previous DC data \cite{collignon2021thermopower}).
As seen in the Hall effect and electrical resistivity of Nd-LSCO~\cite{collignon2017}, the enhancement of $S_{\rm a}/T$ occurs below $T^{\star}$ and reaches a maximum as $T \rightarrow 0$ that corresponds roughly to a 5-fold increase between $p = 0.24$ and $p = 0.20$.
In agreement with Eq. (\ref{eq:behnia_eq}), this roughly matches the change in carrier density from $n \simeq 1 + p = 1.24$ at $p = 0.24$ to $n \simeq p = 0.20$ at $p = 0.20$, suggesting that it is the principal cause for the enhancement of these three transport coefficients ($R_{\rm H}$, $\rho$, $S$) inside the pseudogap phase.

Turning to $S_{\rm c}/T$ below $p^{\star}$, at $p = 0.22$, we see in Fig.~\ref{fig:SoT_0p24_0p22} that it tracks $S_{\rm c}/T$ at $p = 0.24$ down to about 100~K, but then drops upon further cooling to reach negative values below $T \sim 20$~K.
This behavior is seen for all the measured dopings at and below $p^{\star}$ (Fig.~\ref{fig:SoT_doping}(b)), showing that it is a property of the pseudogap phase.
(We stress that the field dependence of $S_{\rm c}$ is very weak, as seen in Fig. S1(b), where the curves taken at $H = 0$ and 16~T essentially overlap, establishing that the sole effect of the field is to suppress superconductivity and reveal $S_{\rm c}/T$ at low temperatures, not to induce the negative $S_{\rm c}/T$.)

As we discuss below, the highly contrasting behavior between $S_{\rm a}/T$ and $S_{\rm c}/T$ below $p^{\star}$ contradicts Eq.~(\ref{eq:behnia_eq}) that predicts an isotropy between the two current directions, and is a puzzle for standard theories of the Seebeck effect, likely carrying significant hints for the underlying nature of the pseudogap phase.
Whatever mechanism causes the drop in carrier density from $n = 1 + p$ to $n = p$ at $p^{\star}$ and the associated Fermi surface transformation observed by ADMR measurements~\cite{fang2020}, it seems to also affect the asymmetry of the band dispersion around the Fermi level in an anisotropic fashion, as attested by the negative $S_{\rm c}/T$.
$S_{\rm c}$ therefore adds to the list of transport properties undergoing a dramatic change upon crossing $p^{\star}$, such as the negative thermal Hall effect~\cite{grissonnanche2019}, attributed to chiral phonons~\cite{grissonnanche2020}, that suddenly appears in the pseudogap phase of Nd-LSCO.

\section{Discussion}

\subsection{$p > p^{\star}$: particle-hole asymmetric energy-dependent scattering rate}

Previous studies of the Seebeck coefficient of cuprates have largely focused on the behavior above $T_c$, revealing that the in-plane $S(T)$ decreases with increasing doping~\cite{obertelli1992} and is negative at high temperature and high doping, as seen for instance in Bi$_{\rm 2}$Sr$_{\rm 2}$CaCu$_{\rm 2}$O$_{\rm 8+x}$ (Bi2212)~\cite{munakata1992}, LSCO~\cite{elizarova2000}, Bi2201~\cite{kondo2005}, and HgBa$_{\rm 2}$CuO$_{\rm 4+\delta}$~\cite{yamamoto2000}.
Nd-LSCO follows this trend as a function of doping at $T$~=~300~K, as showed in Fig.~S2, and as a function of temperature in Fig.~\ref{fig:SoT_0p24_0p22}(a).
In Nd-LSCO above~$p^{\star}$, at $p = 0.24$, $S_{\rm a}(T)$ is slightly negative at 300~K but becomes positive below about $T \sim 250$~K and keeps increasing upon further cooling (Fig.~\ref{fig:SoT_0p24_0p22}(a)).
This behavior contrasts with overdoped Bi2201 where $S_{\rm a}(T)$ is negative
at all temperatures at dopings close to $p^{\star}$~\cite{lizaire2020transport}.

\begin{figure}[t]
\includegraphics[width=0.4\textwidth]{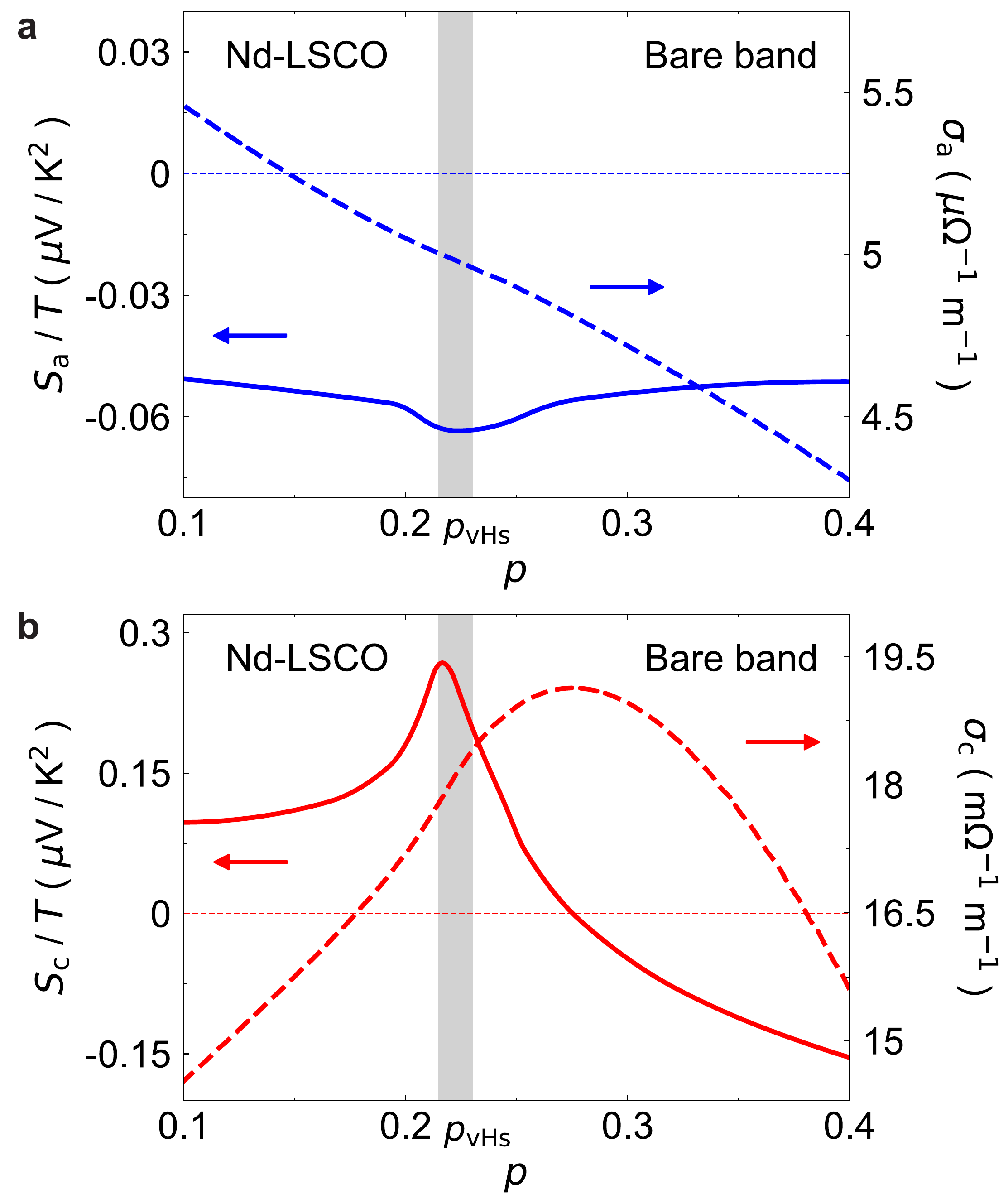}
\caption{Calculations of the Seebeck coefficient ($S/T$, full line, left axis) and conductivity ($\sigma$, dashed line, right axis) as a function of doping from the bare band dispersion of Nd-LSCO, as measured by ARPES \cite{matt2015,horio2018}, with an energy-independent scattering rate $1/\tau$. (\textbf{a}) In-plane coefficients at $T=6$~K; (\textbf{b}) Out-of-plane coefficients at $T=6$~K. We can see from the figures that the sign of $S/T$ is set by the slope of the conductivity $\sigma$ as a function of doping. This reflects the Mott formula in Eq.~(\ref{eq:Mott_formula}). The minimum in $S_{\rm a}/T$ and maximum in $S_{\rm c}/T$ coincide approximately with the van Hove point ($p_{\rm vHs}$, gray band) where the Fermi surface goes from hole-like (below $p_{\rm vHs}$) to electron-like (above $p_{\rm vHs}$).}
\label{fig:calc_vs_p}
\end{figure}

\begin{figure*}[t]
\includegraphics[width=0.65\textwidth]{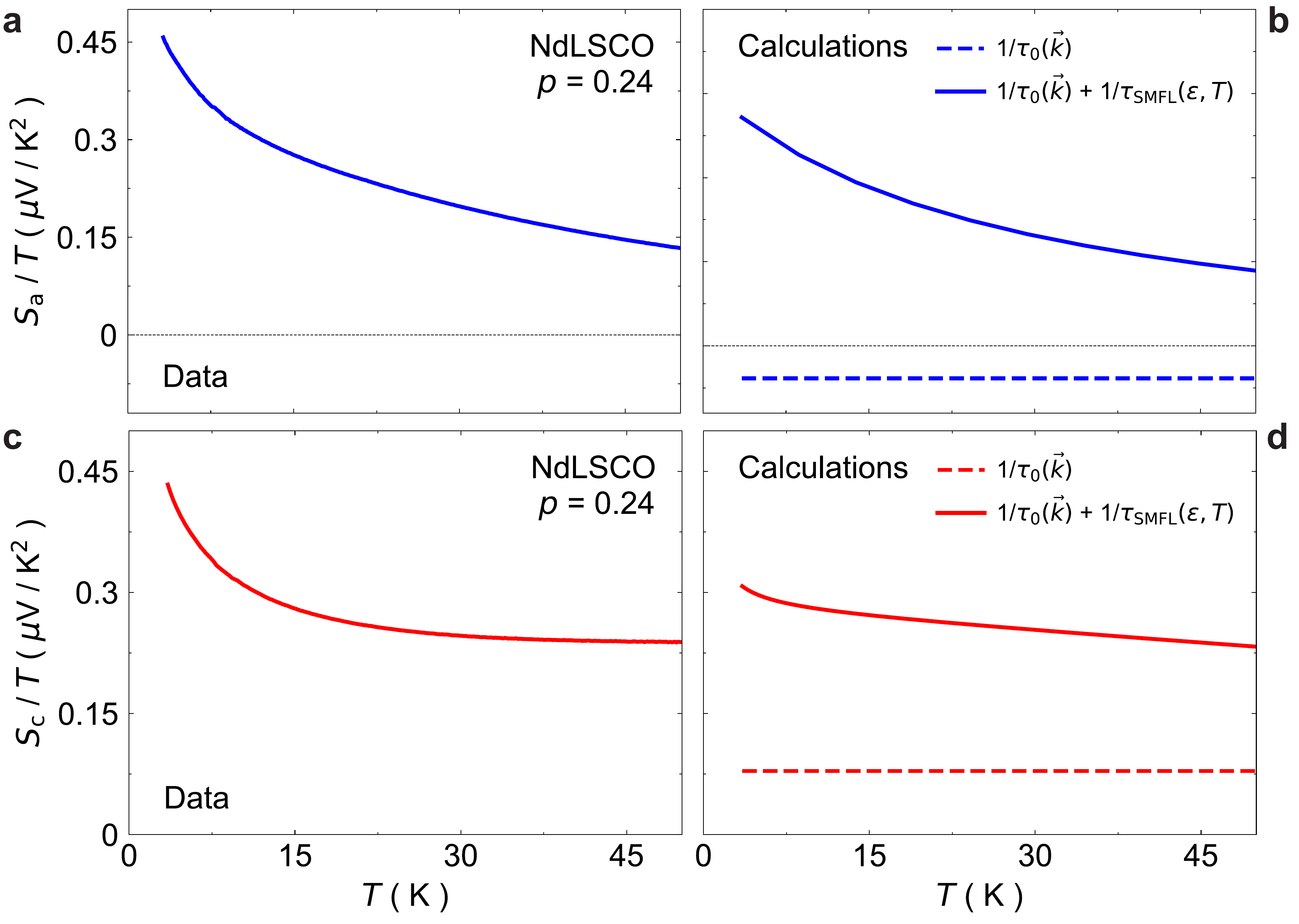}
\caption{Left panels: Seebeck coefficient of Nd-LSCO $p~=~0.24$ for both $a$-axis (blue) and $c$-axis (red). Right panels: Corresponding calculated Seebeck coefficients for a momentum-dependent but energy-independent scattering rate $1/\tau_{\rm 0}(\vecck) = A + B |\cos (2\phi)|^\nu$ (see Appendix~\ref{app:seebeck_calculations} for more details) inferred from the elastic scattering rate extracted from ADMR measurements \cite{grissonnanche2020b} (dashed line) and a momentum- and energy-dependent skewed scattering rate of the form $1/\tau_{\rm 0}(\vecck) + 1/\tau_{\rm SMFL}(\epsilon, T)$ (full line).
}
\label{fig:calc_anisotropic_vs_T}
\end{figure*}

In order to understand the observed behavior of $S_{\rm a}/T$ and $S_{\rm c}/T$, we turn to
Boltzmann transport theory in a relaxation time approximation, using the experimentally determined (quasiparticle) band dispersion $E(\vecck)$
of Nd-LSCO at $p = 0.24$ obtained from ADMR~\cite{fang2020,grissonnanche2020b} and ARPES measurements~\cite{matt2015,horio2018} (see Appendix~\ref{app:seebeck_calculations} for details).
It is certainly not immediately obvious that the Boltzmann formalism can be successfully used to describe the non-Fermi liquid regime considered in the present work.
The justification of this procedure and a discussion of its limitations, including a comparison to a full Kubo calculation,
are provided in Appendix~\ref{app:kubo_vs_boltz}.
We start by considering a constant scattering rate $1/\tau$, independent of energy and momentum.
In this case, the Seebeck coefficient does not depend on $\tau$ because it is a ratio of two transport coefficients, the Peltier coefficient $\alpha$ and the conductivity $\sigma$, and it is given by the Mott formula at low temperature:
\begin{equation}
S_{\rm Mott} = - \frac{\pi^2 k_B^2 T}{3e}\frac{\partial \ln \sigma_{\rm ii}(\epsilon) }{\partial \epsilon}\Big\vert_{\epsilon=0} = - \frac{\pi^2 k_B^2 T}{3e}\frac{\sigma_{\rm ii}^{\prime}(\epsilon)}{\sigma_{\rm ii}(\epsilon)}\Big\vert_{\epsilon=0},
\label{eq:Mott_formula}
\end{equation}
In this expression, $e$ is the electron charge, $k_B$ the Boltzmann constant, $i = x, z$ and
the Fermi level has been set at $\epsilon=0$.
$\sigma(\epsilon)$ is the energy-dependent transport function which only depends on the band structure of the material for a constant $\tau$. Hence, the sign and magnitude of $S$ are entirely determined by the particle-hole asymmetry of the band structure in this case.
In Fig. \ref{fig:calc_vs_p} we display the calculated $S$ for both directions
as a function of doping.
These calculations are performed by allowing for the momentum dependence of the scattering rate along the Fermi surface, inferred from the ADMR measurements \cite{grissonnanche2020b}.
We see that the sign of $S$ is simply set by the slope of $\sigma(\epsilon=0)$ vs $p$ (the inverse of the resistivity $\rho$). In Fig. \ref{fig:calc_vs_p}(a), the in-plane electrical conductivity $\sigma_{\rm a})$ decreases monotonically with doping. This results in a negative $S_{\rm a}$ predicted for all overdoped cuprates. The sign of $S_{\rm a}$ does not change in the considered doping range, whereas the sign of $S_{\rm c}$ flips (becomes negative) at high $p$. One can see that crossing the van Hove singularity around $p = p_{\rm vHs}$ has a large effect on the out-of-plane coefficient, but a modest one for the in-plane coefficient. This is because above $p_{\rm vHs}$ the Fermi surface loses states in the antinodal regions, where the $c$ axis dispersion is largest.

We turn to the temperature dependence of the Seebeck coefficients $S_{\rm a}/T$ and $S_{\rm c}/T$ in Nd-LSCO at $p = 0.24$.
Fig.~\ref{fig:calc_anisotropic_vs_T} shows a direct comparison between our experimental data (left column) and the calculations (right column).
We see from Fig.~\ref{fig:calc_anisotropic_vs_T}(b) and (d), that the calculated $S(T)$ assuming an energy-independent (but momentum-dependent) scattering rate (dashed lines) disagrees with the data in magnitude for both directions.
Furthermore, it yields the wrong sign for the in-plane component, $S_{\rm a}$,
as also found in Ref.~\cite{verret2017}.
To summarize, Fig.~\ref{fig:calc_anisotropic_vs_T} shows that neither the sign of $S_{\rm a}$ nor the $T$ dependence and magnitude of $S_{\rm a}$ and $S_{\rm c}$ can be explained when assuming that particle-like and hole-like excitations have equal scattering rates, even when taking into account the momentum dependence of this scattering rate along the Fermi surface.

We now consider the effect of a scattering rate which depends on energy, allowing for this energy dependence to reflect an asymmetry between particle and holes.
The energy dependence of the inelastic scattering rate has long been recognized to be unconventional in cuprates.
It is often described by the marginal Fermi liquid (MFL) form~\cite{varma1989,varma2020}:
$1/\tau_{\rm MFL} = \sqrt{(a \epsilon)^2 + (\alpha \frac{k_{\rm B} T}{\hbar})^2}$.
As written, this expression is symmetric (even) under $\epsilon\rightarrow - \epsilon$, corresponding to equal scattering rates for particles and holes.
Here, we introduce a modification of the MFL scattering rate with a particle-hole asymmetric energy dependence (`skewed' marginal Fermi liquid - SMFL), namely:
\begin{equation}
1/\tau_{\rm SMFL}(\epsilon, T)=\sqrt{(a_{\pm}\epsilon)^2 + (\alpha \frac{k_{\rm B} T}{\hbar})^2}
\label{eq:SMFL}
\end{equation}
in which $a_{+}$ applies to $\epsilon>0$ and $a_{-}$ to $\epsilon<0$. The particle-hole asymmetry is encoded in the difference between the coefficients $a_{+}$ and $a_{-}$.
We justify the use of the marginal Fermi liquid ansatz for Nd-LSCO at $p=0.24$ by the observation of perfectly $T$-linear resistivity (along both the $a$ axis and the $c$ axis) below $T \simeq 50$~K~\cite{daou2009,legros2019}.
The dimensionless coefficient $\alpha = 1.2 \pm 0.4$ extracted from ADMR~\cite{grissonnanche2020b} obeys the so-called Planckian limit~\cite{bruin2013,legros2019}.
In addition to this SMFL inelastic scattering rate, we also take into account, as above, the momentum-dependent
elastic scattering rate $1/\tau_{\rm 0}(\vecck)$ obtained from the ADMR experiment~\cite{grissonnanche2020b}
for Nd-LSCO at $p=0.24$. Hence, we use in our calculations the total scattering rate:
\begin{equation}
1/\tau(\epsilon, \vecck, T) = 1/\tau_{\rm 0}(\vecck) + 1/\tau_{\rm SMFL}(\epsilon, T).
\label{eq:total_scattering}
\end{equation}
Note that the ADMR study yields the remarkable finding that the inelastic scattering rate is isotropic, \textit{i.e.} $k$-independent. Correspondingly, we assume the energy dependence to also be isotropic, \textit{i.e.} we take $a_+$ and $a_-$ to be $k$-independent.
In Fig.~\ref{fig:calc_anisotropic_vs_T}(b) and (d), we see that the asymmetric scattering rate (plotted in Fig.~S3 at $T=25$~K) changes the calculated $S(T)$ dramatically for both directions and matches well the experimental data.
Crucially, the calculated $S_{\rm a}(T)$ now has the correct sign. Furthermore, $S_{\rm c}(T)$ is reproduced well, including the nontrivial plateau behavior between 15~K and 60~K.
The same model and the exact same parameters lead to calculated in-plane and out-of-plane resistivities that also reproduces the data well in agreement with Ref.~\cite{grissonnanche2020b}.
It is important to note that the particle-hole asymmetry of the scattering rate has a radical effect on the Seebeck coefficient, but no effect on the resistivity.

\begin{figure}[t!]
\includegraphics[width=0.5\textwidth]{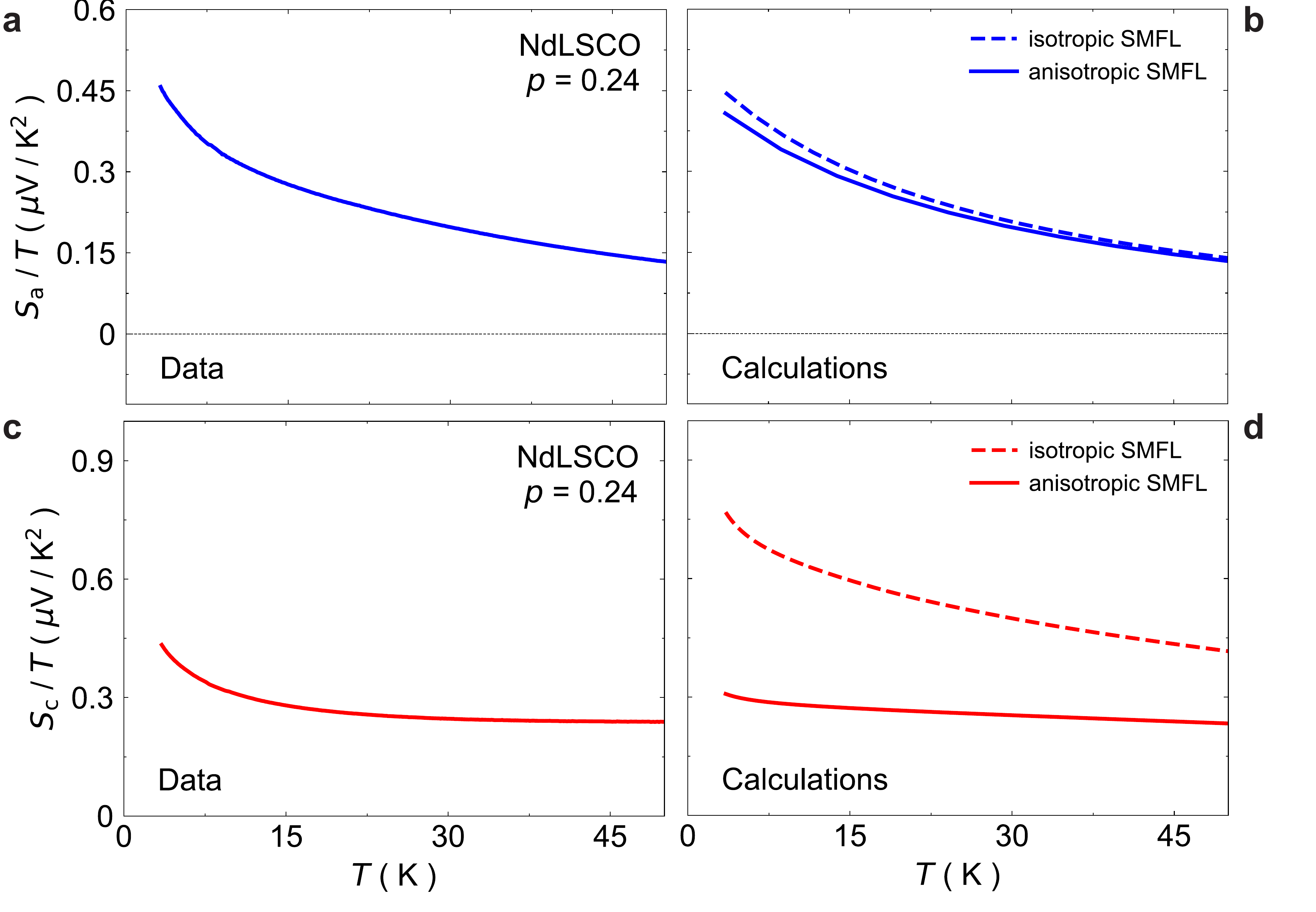}
\caption{Left panels: Seebeck coefficient of Nd-LSCO $p~=~0.24$ for both $a$-axis (blue) and $c$-axis (red). Right panels: Corresponding calculated Seebeck coefficients for a momentum-independent but energy-dependent skewed scattering rate $1/\tau_{\rm 0} + 1/\tau_{\rm SMFL}(\epsilon, T)$, with $B=0$ in Eq.~S5 for the elastic scattering rate (isotropic, dashed line) and a momentum and energy-dependent skewed scattering rate of the form $1/\tau_{\rm 0}(\vecck) + 1/\tau_{\rm SMFL}(\epsilon, T)$ (anisotropic, full line), same as in Fig. \ref{fig:calc_anisotropic_vs_T}.
}
\label{fig:calc_isotropic_vs_T}
\end{figure}

In Fig. \ref{fig:calc_isotropic_vs_T} we reveal the importance of incorporating the anisotropic elastic scattering rate $1/\tau_{\rm 0}(\vecck)$, inferred from the ADMR \cite{grissonnanche2020b}, in the calculations of the Seebeck coefficient.
We see that using an isotropic scattering rate does not affect $S_{\rm a}$ much, it greatly enhances the amplitude of $S_{\rm c}$ and does not exhibit the experimentally observed plateau which was reproduced with an anisotropic scattering rate.
It is not surprising that $S_{\rm c}$ is more affected by a large anisotropic scattering rate, as most of the $c$ axis dispersion is located closer to the antinodal region of the Fermi surface, the same region whose contribution gets dramatically reduced by the anisotropy of the scattering rate.
It therefore appears now that the plateau in $S_{\rm c}/T$ below 60~K (Fig.~\ref{fig:SoT_0p24_0p22})results from a competition between two effects that grow upon cooling: first the skewness increases upon cooling (see Fig.~S3) which increases $S_{\rm c}/T$, second the anisotropy of the scattering rate along the Fermi surface that decreases $S_{\rm c}/T$.

We now provide insight into the temperature-dependence and sign of the Seebeck coefficient
found from this calculation.
The SMFL expression (3) is a particular example of a general class of ‘skewed’ non-Fermi
liquids recently discussed by two of us in Ref.~\cite{georges2021}.
In that work, it is shown that an inelastic scattering rate obeying an $\epsilon/T$ scaling with an intrinsic particle-hole asymmetry of the scaling function remarkably leads to a modification of the low-temperature value of the Seebeck coefficient, even in the presence of elastic scattering.
This is in strong contrast to Fermi liquids, in which the dominant term in the inelastic scattering rate is particle-hole symmetric and particle-hole asymmetric subdominant corrections affect the Seebeck coefficient only at high temperature but not in the low-$T$ regime dominated by elastic scattering~\cite{georges2021,haule_2009}.
In the case of the SMFL, following a simplification that is discussed in the Appendix~\ref{app:kubo_vs_boltz} and detailed in Ref.~\cite{georges2021}, in the $T\rightarrow 0$ limit we obtain:
\begin{equation}
\frac{S}{T}\bigg|^{\mathrm{SMFL}}_{T=0} \simeq
\frac{1}{Z(T)}\left(\frac{S}{T}\right)_{{\rm Mott}} \, +  c_{\rm a}\, \frac{k_B^2}{e \hbar} \alpha\,\tau_{\rm 0}
\label{eq:Seebeck_slope_lowT_Planck}
\end{equation}
In this expression, the Mott value corresponds to expression (\ref{eq:Mott_formula}) which ignores the energy dependence of the scattering rate and whose sign is entirely determined by band structure.
$Z(T)$ is a mass enhancement renormalization to which we return below.
Here, we want to put the emphasis on the second term in (\ref{eq:Seebeck_slope_lowT_Planck}), which
involves a coefficient $c_{\rm a}$ determined  by the particle-hole asymmetry (`skew') of the scattering rate ($c_{\rm a}= \int_0^\infty dx x/[4 \cosh(x/2)^2] [\sqrt{1 + (a_+ \hbar x /\alpha)^2}-\sqrt{1 + (a_- \hbar x /\alpha)^2}]  \approx 1.1 \hbar  (a_+ - a_-) / \alpha \approx 0.4$ for the values of parameters reported in Appendix~\ref{app:seebeck_calculations}), the dimensionless coupling constant $\alpha$ measuring the strength of the inelastic scattering, and the elastic scattering time $\tau_{\rm 0}$.
This unconventional contribution to the Seebeck coefficient implies that the particle-hole asymmetry of the inelastic scattering rate can affect the magnitude and the sign of the Seebeck coefficient even in the low temperature limit where the magnitude of the inelastic scattering is small in comparison with the elastic scattering. This effect applies when the scattering rate is non-Fermi liquid and explains the observed sign change in the calculation above.
We also note that, in contrast to the band (Mott) term, this contribution does not depend on any details of the band structure and is, in particular, isotropic. This may provide a hint into the observed isotropy of our experimental data at low temperature.

%

We now discuss the renormalization $Z(T)$. By Kramers-Kronig, the ansatz for the scattering rate (imaginary part of the self energy) with a linear energy/temperature dependence  implies a logarithmic term in the low-energy slope of the real part of self energy, that is, a mass renormalization
$1/Z(T)=m^*/m \propto (1 + \log(\Lambda/T))$, where $\Lambda$ is a cutoff.
Indeed, a logarithmic enhancement of the specific heat coefficient has been observed experimentally for Nd-LSCO at $p=0.24$~\cite{michon2019}.
We see that this renormalization in turn implies a logarithmic behavior of $S/T$ at low temperatures, as seen in the first term of Eq.~\ref{eq:Seebeck_slope_lowT_Planck}.
It is important to notice however that this renormalization multiplies the band (Mott) part of the Seebeck coefficient.
As a consequence, when the logarithmic mass term is included in our modeling its effect is to increase $S_{\rm c}$ (improving the agreement with the experiment), but also to
diminish $S_{\rm a}$ (worsening the agreement somewhat for this direction). A further understanding of the mass renormalization effect is left for future work.
Note that logarithmic terms in the $T$-dependence of the Seebeck coefficient have been
discussed in \cite{paul_kotliar_2001} in relation to the proximity of a quantum critical point.

Recently, Jin \textit{et al.}~\cite{jin2021} reported measurements of in-plane Seebeck coefficient for LSCO $p=0.33$. At this doping, $S_{\rm a}$ displays also a positive sign down to the lowest temperature, while it shows a $T^2$ resistivity characteristic of a Fermi liquid. Because particle-hole asymmetry in a Fermi liquid cannot cause a change of sign in the low-temperature regime \cite{georges2021}, this could point to the influence of non-Fermi liquid corrections at the antinodes even in the overdoped regime~\cite{chang2013}.

We conclude this section by contrasting our analysis with previous calculations of the Seebeck coefficient of other families of cuprate compounds for $p>p^\star$.
In overdoped Bi2201 and Bi2212, in which the in-plane Seebeck coefficient is mostly negative, calculations by Kondo et al.~\cite{kondo2005} successfully reproduced the $T$-dependence of $S$ by using the electronic structure measured by ARPES and assuming that the scattering time depends on momentum in such a way as to maintain a constant scattering length $l= v \tau$, that is  $\tau(\vecck,\epsilon) \propto 1 / v(\vecck,\epsilon)$ (the same assumption was made in Ref.~\cite{storey2013}).
In this expression, $\vecck$ denotes a momentum on the Fermi surface and $\epsilon$ the energy of an excitation when moving away from the Fermi surface. The asymmetry between $\epsilon>0$ (particles) and $\epsilon<0$ (holes) in this model originates from the nearby van Hove singularity for momenta near the antinodes.
Although this model was able to capture the temperature dependence and sign of the data, its validity is uncertain.
Indeed, in this model the scattering rate is the largest at the nodes, where the velocity is maximum. This is at odds with most of the literature on cuprates \cite{abdeljawad2006, chang2013, grissonnanche2020b,yoshida_2007}.

Finally, it is interesting to note that, although quasi-1D organic superconductors can have very different physics from cuprates at the pseudogap critical point, calculations on a microscopic model have shown that incorporating the energy dependence of the scattering rate in those materials causes the Seebeck coefficient to change sign when close to the magnetic quantum critical point \cite{shahbazi2016}.

\subsection{$p < p^{\star}$: Fermi surface transformation}
In Nd-LSCO, reducing the doping below $p^{\star}$ induces two main anomalies in the Seebeck coefficient at low temperatures: a large enhancement of $S_{\rm a}$ and a negative $S_{\rm c}$.
Both are unambiguously connected with the onset of the pseudogap phase: they occur upon crossing $p^{\star}$, and at low temperatures below the pseudogap temperature $T^{\star}$.
The sudden sign change in the out-of-plane Seebeck coefficient necessarily reflects a change in particle-hole asymmetry upon entering the pseudogap phase, either through a transformation of the electronic structure, and therefore the Fermi surface, or the scattering rate.
As discussed previously~\cite{collignon2021thermopower}, the enhancement in $S_{\rm a}$ below $p^{\star}$ reflects the change in $n$ from $n \simeq 1 + p$ above $p^{\star}$ to $n \simeq p$ below $p^\star$ associated with the Fermi surface transformation at $p^{\star}$~\cite{fang2020}.
The origin of the negative $S_{\rm c}$, however, is not clear and is the focus of the remaining discussion.

In Nd-LSCO, the Fermi surface undergoes two changes at $p^{\star}$.
First, the large Fermi surface goes from hole-like to electron-like when the van Hove singularity crosses the Fermi level, which occurs at a doping $p_{\rm vHs}$ that coincides with $p^{\star}$~\cite{matt2015}.
As we show in Fig. \ref{fig:calc_vs_p}, the bare band calculations predict no sign change in $S_{\rm a}$ across $p_{\rm vHs}$, and a change to a positive sign in $S_{\rm c}$, in contrast to our results.
In Fig.~SS2, we plot $S_{\rm a}/T$ and $S_{\rm c}/T$ at 300~K as a function of doping and see that both quantities decrease smoothly from $p = 0.20$ to $p = 0.24$, with no anomaly.
Similarly, we observe that $S_{\rm a}/T$ and $S_{\rm c}/T$ show the same smooth increase upon cooling from 300~K down to about 100~K (Figs.~\ref{fig:SoT_doping}), irrespective of doping.
At those high temperatures above $T^{\star}$, only the vHs has an impact on the band structure, and since there is no clear difference in the behaviors of $S_{\rm a}/T$ and $S_{\rm c}/T$ we conclude that the vHs alone is not the main cause for the negative $S_{\rm c}/T$.
Second, there is a transformation of the Fermi surface associated with the drop in carrier density as deduced from Hall effect, resistivity~\cite{collignon2017}, thermal conductivity~\cite{michon2018}, and in-plane Seebeck~\cite{collignon2021thermopower} measurements.
This Fermi surface transformation was recently detected directly by ADMR measurements~\cite{fang2020}, which found that the Fermi surface below $p^{\star}$ is likely made of four nodal hole pockets, consistent with a transformation by a $Q = (\pi,\pi)$ wavector.
ARPES measurements see Fermi arcs that are consistent with the side of these pockets residing in the 1st Brillouin zone \cite{matt2015}. A model is needed to explain how such a Fermi surface transformation can account for a negative $S_{\rm c}$.

The scattering rate, on the other hand, is not known to undergo a significant change at $p^{\star}$.
Resistivity measurements on either side of $p^{\star}$~\cite{collignon2017} find that the normal-state magnetoresistance (MR) is comparable at $p = 0.22$ and $p = 0.24$.
In the weak-field limit, the magnetoresistance varies as MR~$\propto(\omega_c \tau)^2$, where $\omega_c$ is the cyclotron frequency and $\tau$ is the scattering time, suggesting that $\tau$ does not change significantly across $p^{\star}$.
Similarly, in ADMR measurements on Nd-LSCO~\cite{fang2020} the magnitude of the oscillations seen on each side of $p^{\star}$, at $p = 0.21$ and $p = 0.24$, is similar, implying that the amplitude of $\tau$ is roughly the same.
So there is no indication, a priori, that the sign change in $S_{\rm c}$ is correlated with a change in $\tau$ at the Fermi level.
Finally, the fact that the change of $S_{\rm a}$ through $p^{\star}$ roughly matches the ratio of the carrier densities further indicates that any change in $\tau$ is small.

A change in Fermi surface, and specifically in the asymmetry of the band near the Fermi level is therefore the likely cause of the negative $S_{\rm c}$.
It seems clear that without the pseudogap, which appears at temperatures below $\sim 100$~K over this doping range, $S_{\rm a}/T$ and $S_{\rm c}/T$ would keep evolving in tandem down to low temperatures, likely to match in the $T \rightarrow 0$ limit as they do at $p = 0.24$. (This was actually shown to be the case by application of pressure to move $p^\star$ down below $p = 0.22$: the low-$T$ rise in $S_{\rm a}/T$ at $p = 0.22$ seen at ambiant pressure is gone under pressure~\cite{gourgout2021pressure}.)
Consequently, we attribute the negative $S_{\rm c}$ to a transformation of the band structure and therefore of the Fermi surface caused by the pseudogap phase.
Based on ARPES~\cite{matt2015} and recent ADMR measurements~\cite{fang2020}, the Fermi surface below $p^{\star}$ is truncated at the antinodes. Now this antinodal region is where $c$ axis dispersion is largest, thereby dominating, lending the bulk of the $c$ axis transport. From there, it is not hard to imagine that out-of-plane Seebeck would be most sensitive to a transformation of the Fermi surface in the ($\pi$, 0) direction caused by the pseudogap phase. Further work  is needed to model the sign change of $S_{\rm c}$ below $p^*$.

\section{Summary}

In summary, we have measured the Seebeck coefficient of the cuprate superconductor Nd-LSCO at dopings close to the critical doping \pstar~$= 0.23$ where the pseudogap phase ends at $T=0$.
In particular, we examine the $c$ axis component, $S_{\rm c}$, for a heat current normal to the CuO$_2$ planes, in the presence of a magnetic field sufficient to suppress superconductivity and thus track the normal-state behaviour down to low temperature.

At $p >$~\pstar, for $p = 0.24$, we find that $S_{\rm c}(T)$ and the in-plane component $S_{\rm a}(T)$ are both positive below 250~K, and become roughly equal below 10~K. This is in contrast to what calculations based on the well-characterized band structure predict, namely that $S_{\rm a}$ should be negative at all temperatures. We show that a good quantitative description of (and the correct sign for) both $S_{\rm c}(T)$ and $S_{\rm a}(T)$ is obtained if we add a linear asymmetric energy dependence to the scattering rate $1/\tau$ previously extracted in Nd-LSCO as a function of angle and temperature via ADMR measurements~\cite{grissonnanche2020b}. 
This suggests that $1/\tau$ is Planckian not only in its $T$ dependence ($1/\tau \simeq k_{\rm B}T / \hbar$) but also in its energy dependence ($1/\tau \propto \epsilon$),
with an intrinsic particle-hole asymmetry of the $\epsilon/T$ scaling function -- as expected for a skewed Planckian metal~\cite{georges2021}.

At $p <$~\pstar, we find that $S_{\rm c}(T)$ undergoes a dramatic change at low temperature, suddenly becoming negative as soon as $p$ falls below \pstar. This is a striking new experimental signature of the pseudogap phase, which we attribute to a transformation of the Fermi surface. Further work is needed to identify what modification of the Fermi surface is responsible for this sign change.

\section*{Acknowledgements}

We thank S.~Fortier for his assistance with the experiments.
L.T. acknowledges support from the Canadian Institute for Advanced Research
(CIFAR) as a CIFAR Fellow
and funding from
the Institut Quantique,
the Natural Sciences and Engineering Research Council of Canada (PIN:123817),
the Fonds de Recherche du Qu\'ebec -- Nature et Technologies (FRQNT),
the Canada Foundation for Innovation (CFI),
and a Canada Research Chair.
J.M. is supported by the Slovenian Research Agency (ARRS) under Program No. P1-0044 and Project No. J1-2458, N1-0088, and J1-2455-1.
J.S.Z. was supported by NSF MRSEC under Cooperative Agreement No. DMR-1720595.
This research was undertaken thanks in part to funding from the Canada First Research Excellence Fund
and the Gordon and Betty Moore Foundation's EPiQS Initiative (Grant GBMF5306 to L.T.).
The Flatiron Institute is a division of the Simons Foundation.


\appendix


\section{Field dependence of $S$}
\label{app:field_dependence}

\begin{figure}[h!]
\includegraphics[width=0.42\textwidth]{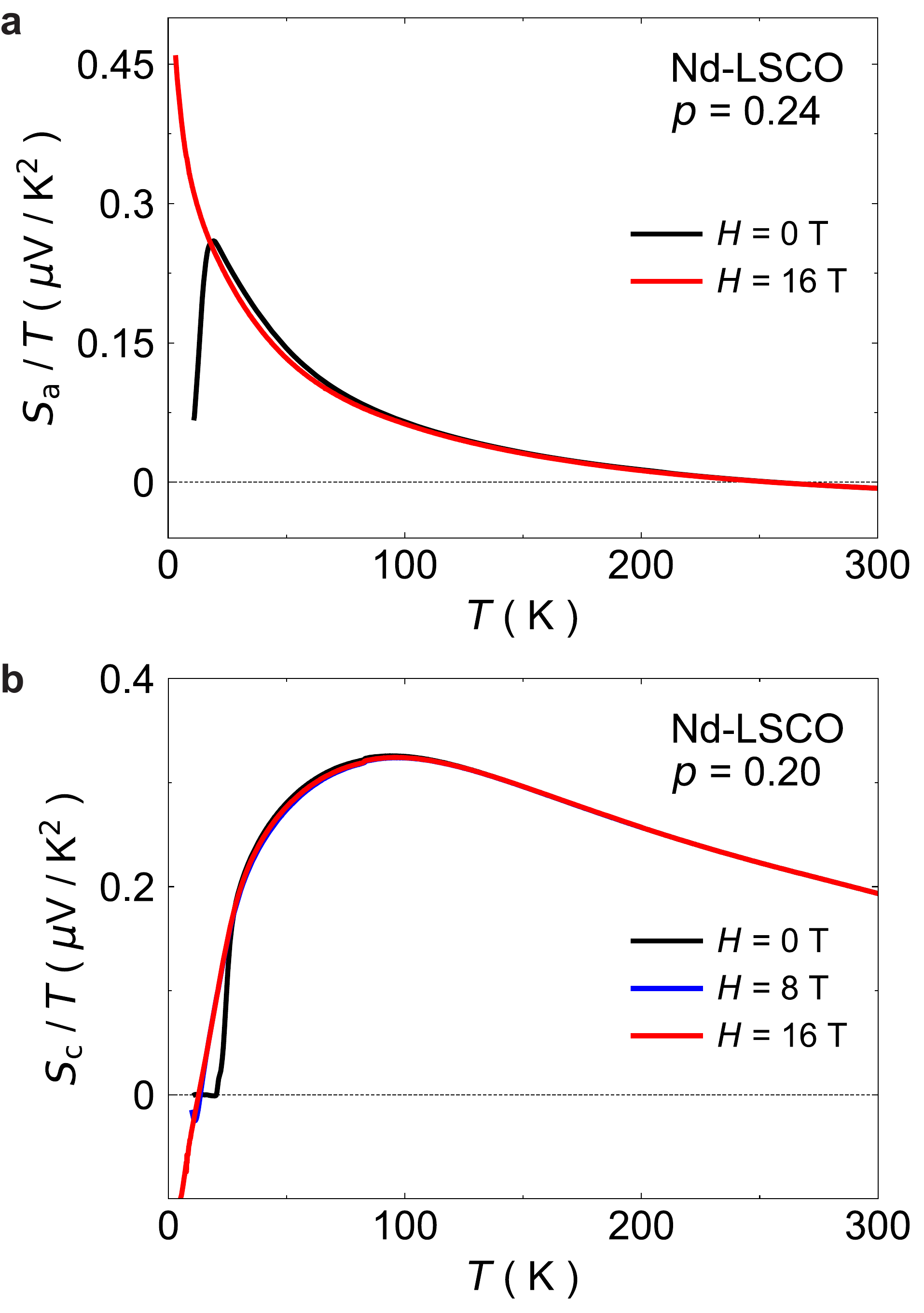}
\caption{$S/T$ of Nd-LSCO as a function of temperature for different fields applied parallel to the $c$ axis. (\textbf{a}) $S_{\rm a}/T$ at $p = 0.24$ in $H=0$ and $16$~T. (\textbf{b}) $S_{\rm c}/T$ at $p = 0.20$ in $H=0$, $8$ and $16$~T. We see that applying a field has very little effect on the normal-state $S_{\rm c}/T$ curve, except for suppressing superconductivity and revealing it down to low temperatures.}
\label{fig:field_dependence}
\end{figure}

\section{Doping dependence of $S/T$ at $T= 300$~K}
\label{app:doping_dependence}

\begin{figure}[h!]
\includegraphics[width=0.45\textwidth]{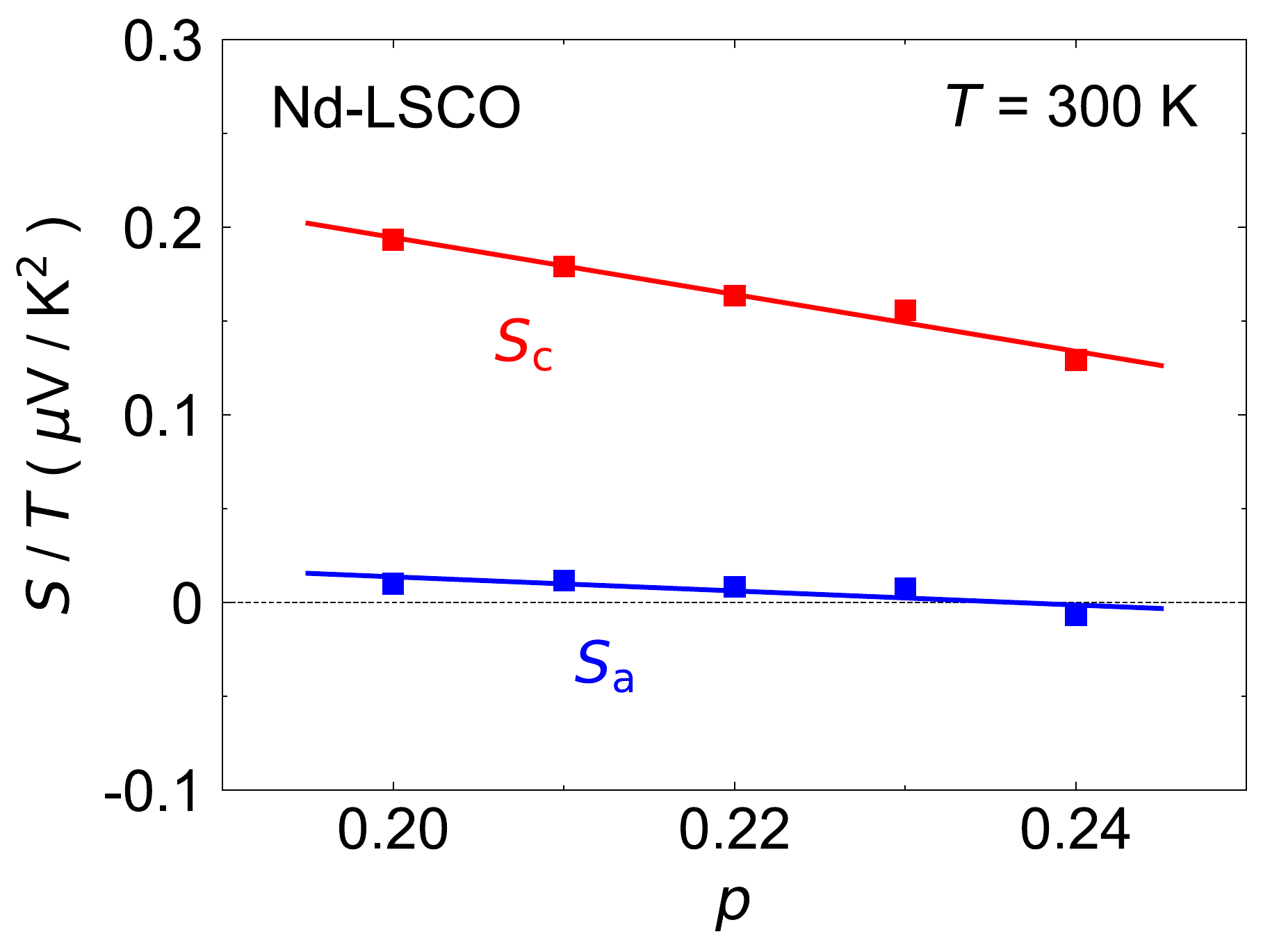}
\caption{$S_{\rm a}/T$ (blue) and $S_{\rm c}/T$ (red) at $T=300$~K (and $H=0$) as a function of doping for Nd-LSCO $p$ = 0.20, 0.21, 0.22, 0.23 and 0.24. Solid lines are a linear fit to the data.}
\label{fig:doping_dependence_300K}
\end{figure}

Like in previous studies of the Seebeck coefficient of cuprates, the in-plane $S(T)$ goes down with increasing doping~\cite{obertelli1992,munakata1992,elizarova2000,kondo2005,yamamoto2000} and is negative at high temperature and high doping.
The in-plane Seebeck coefficients of Nd-LSCO also follows this trend as a function of doping at $T$~=~300~K, as showed in Fig.~\ref{fig:doping_dependence_300K}. The out-of-plane Seebeck coefficient at $T=300$~K also decreases with doping.

\section{Seebeck calculations}
\label{app:seebeck_calculations}

\begin{figure}[b!]
\includegraphics[width=0.4\textwidth]{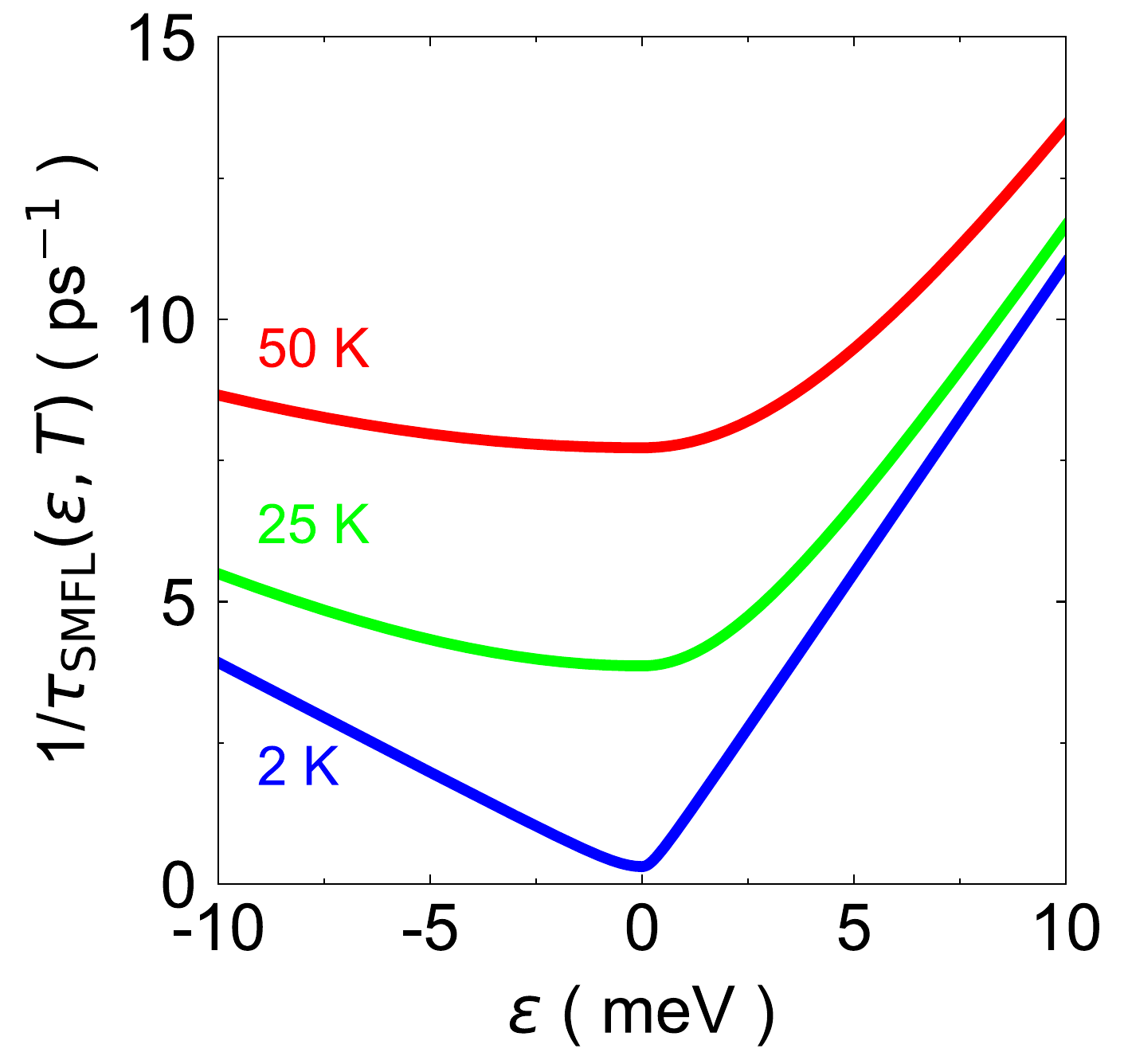}
\caption{Energy dependence of the skewed marginal Fermi liquid scattering rate $1/\tau_{\rm SMFL}~(\epsilon, T)$ at $T=2$, 25 and 50~K. $1/\tau_{\rm SMFL}$ is defined by Eq.~(\ref{eq:SMFL}) with parameters given in Table~\ref{tab:0p24_scattering}.}
\label{fig:scattering_vs_energy}
\end{figure}

The Seebeck coefficient is given by the ratio of the Peltier coefficient $\alpha_{ii}$ to the electrical conductivity $\sigma_{ii}$ (with $i = x, z$), $S_i =  \alpha_{ii}/\sigma_{ii}$, where
\begin{equation}
\sigma_{\rm ii}
=
\int_{-\infty}^{\infty} d\epsilon
\Big(-\frac{\partial f(\epsilon)}{\partial \epsilon}\Big)\sigma_{\rm ii}(\epsilon)
\label{sigma}
\end{equation}
\begin{equation}
\alpha_{\rm ii}
=
\int_{-\infty}^{\infty}d\epsilon
\bigg[\Big(-\frac{\partial f(\epsilon)}{\partial \epsilon}\Big)\frac{\epsilon}{T}\bigg]\frac{\sigma_{\rm ii}(\epsilon)}{-e}
\label{alpha}
\end{equation}
with $e$ the electron charge, $f(\epsilon)$ the Fermi-Dirac distribution and
\begin{equation}
\sigma_{\rm ii}(\epsilon)
=
2e^2\iiint_{\text{BZ}}\frac{d^3k}{(2\pi)^3}
v_{\rm i}(\vecck)^2
\tau(\vecck, \epsilon)\delta\big(\epsilon-E(\vecck)\big),
\label{eq_will_return}
\end{equation}
where $v_{\rm i}(\vecck)$ is the component of the quasiparticle velocity in the $i$-direction, $\tau(\vecck, \epsilon)$ is the quasiparticle lifetime depending on both momentum $\vecck$ and energy $\epsilon$, and $E(\vecck)$ is the tight-binding band dispersion of Nd-LSCO~\cite{grissonnanche2020b,horio2018}.

In order to calculate Seebeck, we used the tight-binding model $E(\vec k)$, measured by ADMR \cite{grissonnanche2020b} and ARPES \cite{horio2018}, to describe the band dispersion

\begin{equation}
  \begin{aligned}
E(\vec k)=&-2t[\cos(k_xa)+\cos(k_ya)]
\\&-4t'\cos(k_xa)\cos(k_ya)
\\&-2t''[\cos(2k_xa)+\cos(2k_ya)]
\\&-2t_z\pi_{\rm x}\pi_{\rm y}\pi_{\rm z}[\cos(k_xa)-\cos(k_ya)]^2,
  \end{aligned}
    \label{eq:band_structure_0p24}
\end{equation}
with $a=3.75~\mathring{A}$ and $c=13.2~\mathring{A}$ the lattice constants, $\pi_{\rm i}=\cos(k_ia/2)$ with $i=x$, $y$ and $\pi_{\rm z}=\cos(k_zc/2)$. The hopping parameters are found in the following table (from  Extended Data Table I of \cite{grissonnanche2020b})

\begin{table}[h!]
  \begin{center}
    \begin{tabular}{|c|c|c|c|c|c|}
    \hline
     $t$ (meV) & $t'$ & $t''$ & $t_z$ & $\mu$ & $p$\\
    \hline
    160 & $-0.1364t$ & $0.0682t$ & $0.0651t$ & $-0.8414t$ & 0.259\\

    \hline
    \end{tabular}
  \end{center}
 \caption{Tight-binding parameters from the ADMR~data of Nd-LSCO $p$ = 0.24 obtained from the angle-dependent magnetoresistance analysis in \cite{grissonnanche2020b}. The hopping parameter $t=160$~meV was adjusted to meet the specific heat data on the same sample, more details can be found in the Methods section of [16].}
    \label{tab:0p24_bandstructure}
\end{table}

The angle-dependent elastic scattering rate is given by
\begin{align}
1/\tau_{\rm 0}(\vecck) = A + B |\cos (2\phi)|^\nu.
\label{eq:elastic_scattering}
\end{align}
This model was justified by the analysis of the ADMR data in \cite{grissonnanche2020b}. Other functions with more parameters were also employed for comparison and all ended up with the same form factor for elastic scattering rate (see the Method section of \cite{grissonnanche2020b} for more details).

\begin{table}[h!]
  \begin{center}
    \begin{tabular}{|c|c|c|c|c|c|}
    \hline
     \makecell{$a_{\rm -}$ \\(ps$^{-1}$/meV)} & \makecell{$a_{\rm +}$\\ (ps$^{-1}$/meV)} & $\alpha$ & $A$ (ps$^{-1}$) & $B$ (ps$^{-1}$) & $\nu$\\
    \hline
    0.4 & 1.1 & 1.2 & 9.97 & 71.1 & 12\\

    \hline
    \end{tabular}
  \end{center}
 \caption{Scattering rate parameters used in Eq.~(\ref{eq:total_scattering}) and Eq.~(\ref{eq:elastic_scattering}) to calculate the Seebeck coefficients in Fig.~\ref{fig:calc_anisotropic_vs_T}. The value of $\nu$ has been derived from the angle-dependent magnetoresistance and is explained in more details in the Methods section of \cite{grissonnanche2020b}.}
    \label{tab:0p24_scattering}
\end{table}


\section{Comparison between Kubo and Boltzmann results}
\label{app:kubo_vs_boltz}

The use of a Boltzmann formalism in the relaxation time approximation to describe non-Fermi liquid regimes deserves further discussion and justification, which is the subject of this section.
First, it is important to realize that one can derive Boltzmann-like expressions as given by Eqs.~\ref{sigma},\ref{alpha},\ref{eq_will_return} starting from the Kubo formula, without assuming the scattering rate to be of the Fermi liquid form~\cite{deng_2013,xu_haule_2013,georges2021}. We neglect vertex corrections (we leave investigation of effects of those for future work) and assume for simplicity
a single electronic band, but otherwise keep the discussion general.
The Kubo expression for the transport coefficients $L_n$ is
\begin{equation}
\label{ln}
L_n = (2 \pi) \int d\omega \omega^n \left(-\frac{df}{d\omega}\right) \sum_\veck (v^\alpha_{0\veck}
)^2 A_\veck^2
\end{equation}
Note that, in contrast to the Boltzmann formalism, this expression involves the {\it bare} (band structure) velocity denoted by
$v^\alpha_{0\veck}$ and the full electronic spectral function. The latter is given by:
\begin{equation}
    A_\veck(\omega)=-\frac{1}{\pi} \mathrm{Im} \frac{1}{\omega-\varepsilon_\veck - \Sigma_\veck(\omega)}
\end{equation}
with $\varepsilon_\veck$  the band energy (shifted by the chemical potential) and
$\Sigma_\veck (\omega)$ the self energy  of an electron with momentum $\veck$ and excitation energy $\omega$.
This expression simplifies to the Boltzmann-like formalism when $A_\veck(\omega)^2$ can be approximated as
$Z^2\,\delta(\omega-\omega_\veck) \tau_\veck/\pi$ in which $\omega_\veck$ is the quasiparticle energy
at which the spectral function is peaked, given by the solution of $\omega-\varepsilon_\veck-\mathrm{Re} \Sigma_\veck(\omega)=0$. $Z$ is the spectral weight carried by this peak and
$1/\tau_\veck= -2 Z \mathrm{Im} \Sigma_\veck(\omega_\veck)$ is the inverse quasiparticle transport  lifetime.
Introducing the renormalized quasiparticle velocity
$v^\alpha_{\veck}=Z v^\alpha_{0\veck}$ one finally obtains the Boltzmann form:
\begin{equation}
    L_n = 2 \int \frac{d^3k}{(2 \pi)^3}      \omega_\veck^n  \left. \left(-\frac{df}{d\omega}\right)\right|_{\omega=\omega_\veck} \left(v_\veck^\alpha\right)^2 \tau_\veck,
    \label{boltz11}
\end{equation}
In a Fermi liquid, quasiparticle excitations dispersing as $\omega_\veck$ have a very long lifetime
$\tau_\veck\sim 1/T^2$ at low-$T$, and a finite $Z$, hence the quasiparticle peak is very narrow and the above replacement leading to the Boltzmann approximation is justified.
However, Landau Fermi liquid behavior is not a necessary condition, and the Boltzmann approximation can apply even
under milder conditions.
In Ref.~\cite{xu_haule_2013}, it was shown for example that a sufficient condition is that the
peak in the spectral function is narrower than $T$.
We also note that in a non-Fermi liquid $Z$ remains non-zero at finite $T$, even when it
vanishes asymptotically as $T\rightarrow 0$.
An alternative derivation was given in~\cite{georges2021,deng_2013} which assumed however a momentum-independent scattering rate.
However, the justification given in these two articles do not apply in the present case since
we include impurity scattering which
dominates at low-$T$ (hence the spectral function peak is not narrower than $T$) and furthermore
this scattering is momentum dependent.
Hence we must proceed differently to justify the Boltzmann approximation.

\subsubsection{Derivation of Boltzmann transport from Kubo formula for momentum dependent scattering }
\label{sss:fs}
We assume the transport to be metallic, i.e. dominated by contributions close to the Fermi surface
(this means that transport is associated with electrons within a sufficiently thin shell in momentum space, but does not imply Fermi-liquid properties of the corresponding states).
We reformulate the momentum integral in Eq.~\ref{ln} by introducing equienergy $\varepsilon_\veck+\mathrm{Re}\Sigma_\veck=\mathrm{const}=\omega$ surfaces in momentum space. We introduce a new set of coordinates $S_k, k_\perp$ where $k_\perp$ denotes the distance from the equienergy surface at $\omega$ and $S_k$ denotes the position on that surface.

\begin{figure}[h!]
\includegraphics[width=0.35\textwidth]{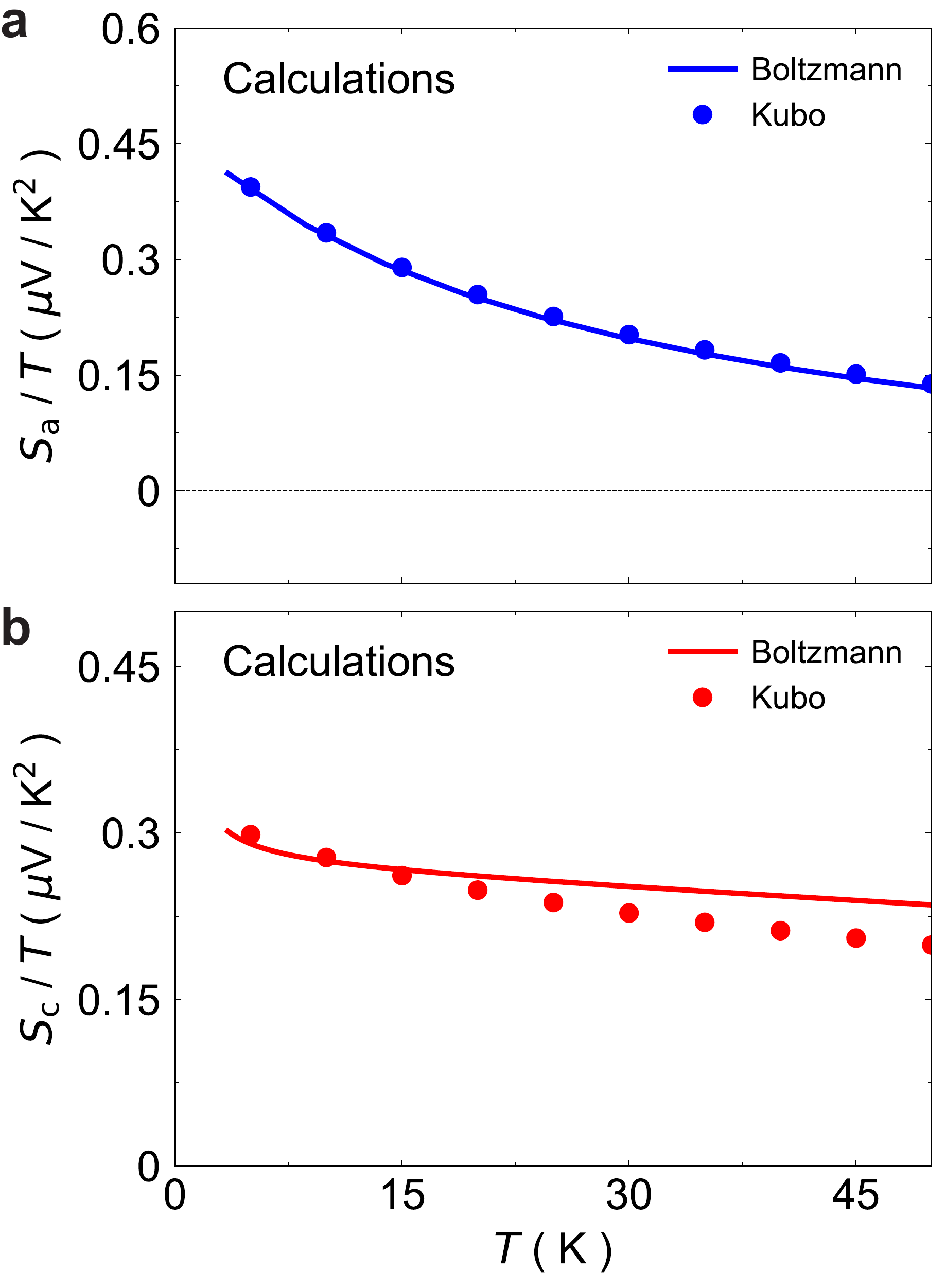} \\
\caption{Comparison between the Kubo and Boltzmann evaluations of the Seebeck coefficient: (a) in-plane $S_{\rm a}$; (b) out-of-plane $S_{\rm c}$.}
\label{fig:kubo_vs_boltz}
\end{figure}

That is, we write
\begin{equation}
    L_n = 2\pi \int d S_k  \int d \omega  \omega^n  \left(-\frac{df}{d\omega}\right)  \int d k  (v_{0\,k_\perp,S_k}^\alpha)^2 A_{k_\perp,S_k}^2 J,
    \label{eq:ln_perp}
\end{equation}
where $J=J(k_\perp,S_k)$ is the Jacobian.
At each position at the Fermi surface $S_k$ and for each $\omega$, the spectral functions peaks at a perpendicular momentum $k_\perp = k_\omega$ given by solution of the equation
$\omega-\varepsilon_{k_\perp,S_k} -\mathrm{Re} \Sigma_{k_\perp,S_k}(\omega)=0$. Linearizing in $k_\perp$ around $k_\omega$, introducing the corresponding "Fermi velocity"
\begin{equation} \tilde{v} =\left.\left[\partial(\varepsilon_{k_\perp,S_k} + \Sigma_{k_\perp,S_k})/\partial k_\perp\right]\right|_{k_\perp =k_\omega}
\end{equation}
and approximating the values of the band velocities and the Jacobian  in the integral over $k_\perp$ in Eq.~\ref{eq:ln_perp} by the value at $k_\omega$ the integrand there becomes the square of Lorentzian function. Extending the integral over $k_\perp$ to $(-\infty,\infty)$ and using $\int dx/(1+x^2)^2  = \pi/2$ one gets
\begin{equation}
L_n = 2 \int d S_k  \int d \omega  \omega^n  \left(-\frac{df}{d\omega}\right) \frac{ \left(v_{0\,k_\omega,S_k}^\alpha\right)^2 }{\tilde{v} (-2 \mathrm{Im} \Sigma_{k_\omega, S_k})}J.
\end{equation}
One can, by reintroducing a momentum variable $k_\perp$ set by relation $\omega =\omega(k_\perp)$ and using $d\omega= \tilde{v} d k_\perp +( \partial \mathrm{Re} \Sigma (\omega)/\partial \omega) d \omega$ and $Z=1-\partial \mathrm{Re} \Sigma / \partial \omega$ rewrite this expression  to
\begin{equation*}
    L_n = 2 \int \frac{d^3k}{(2 \pi)^3}      \omega_\veck^n  \left. \left(-\frac{df}{d\omega}\right)\right|_{\omega=\omega_\veck} \left(v_{\veck}^\alpha\right)^2 \tau_\veck,
    \
\end{equation*}
which is the Boltzmann form Eq.~\ref{boltz11}. Notice that quasiparticle velocities and lifetimes appear in this expression.

Approximating $J, (v^\alpha_{k,S_k})^2 $ and $\mathrm{Im} \Sigma_{k,S_k}$ by the value at $k_\omega$ and taking them out of of the $k_\perp$ integral in  Eq.~\ref{eq:ln_perp} is valid when these quantities vary slowly in $k$, a condition that can be specified in terms of
\begin{equation}
     \frac{1}{v_{F}}\left(\frac{d \log {\zeta}} {dk_\perp}\right)  (-\mathrm{Im} \Sigma) \ll 1 \label{eq:cond}
\end{equation}
for quantities $\zeta=J, v^\alpha_{k_\perp,S_k}$, -$\mathrm{Im} \Sigma_{k_\perp,S_k}$ and that needs to hold for all $S_k$. Crucial for the discussion that follows is the appearance of the Fermi velocity $v_{\omega,S_k} \sim v_F$. When this is small (such as close to van-Hove singularities), this condition becomes more difficult to satisfy. Likewise, close to the boundaries of the Brillouin zone also extending the boundaries of momentum integral becomes problematic.

\subsubsection{Numerical verification}
For a more complete confirmation of the validity of the Boltzmann approach in the present case, hence we investigated the issue numerically. In order to setup a Kubo calculation, we constructed a self-energy with the imaginary part given by (half) of the scattering rate extracted from ADMR experiments and used in the Boltzmann calculations.
Because the renormalizations are already included in the extracted tight-binding parameters, we neglected the real part of self energy. The results are shown in Fig~\ref{fig:kubo_vs_boltz}. One sees that the Boltzmann description captures the results obtained using the Kubo formula quite well. In particular, for the inplane response the calculated Seebeck coefficient lies on top of the Boltzmann description. For the out-of-plane the deviations are larger but remain $\lesssim 20\%$, which shows that there is no qualitative breakdown of the Boltzmann description even though the energy and temperature dependence of the considered scattering are non-Fermi liquid and strongly angular dependent.
This is a direct demonstration of the validity of the Boltzmann-like description used in the main text.

\begin{figure}[b!]
\includegraphics[width=0.4\textwidth]{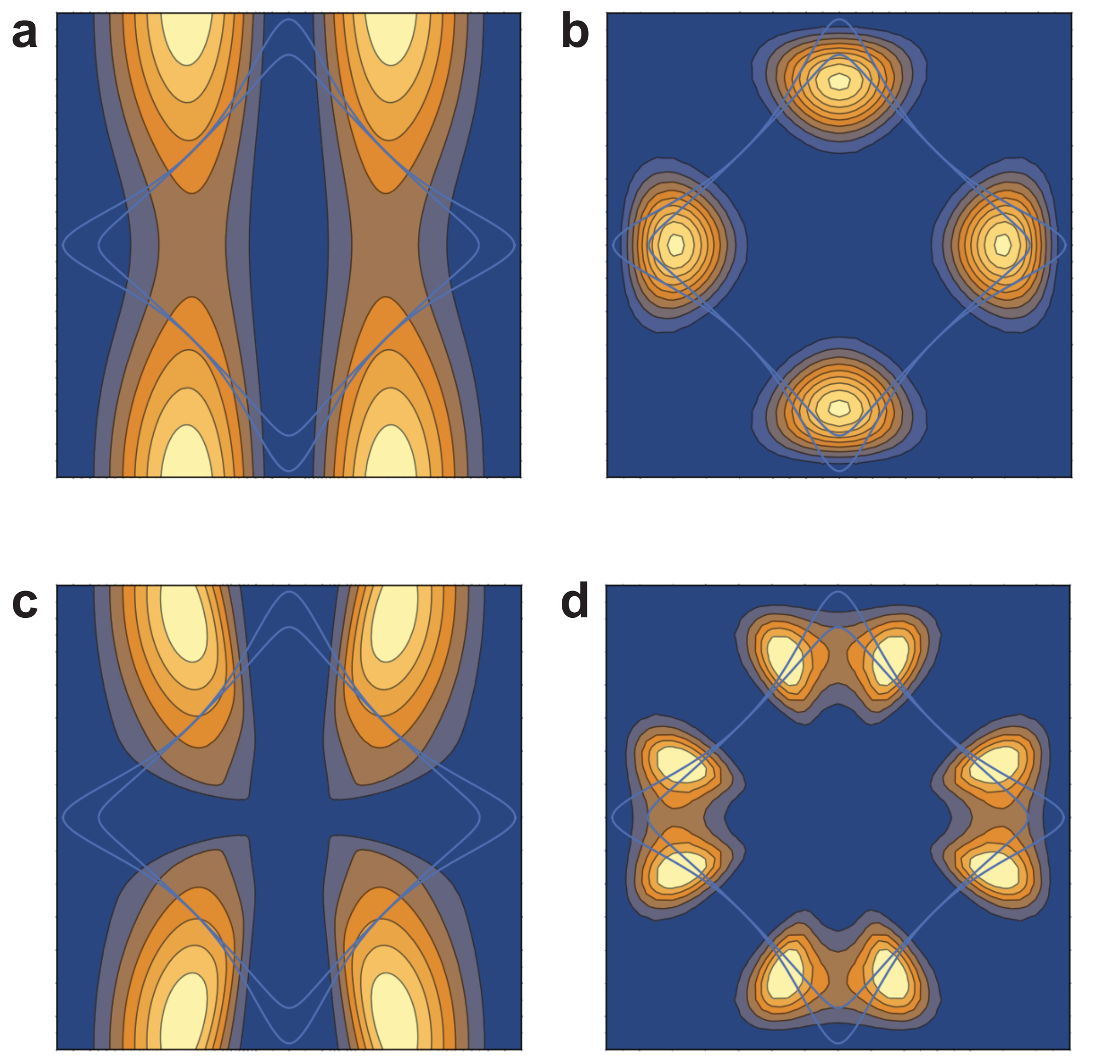}
\caption{Color maps of (a) $v_x^2$, (b) $v_z^2$, (c) $v_x^2 \tau$, and $v_z^2 \tau$ in the Brillouin zone. The Fermi surfaces at $k_z=0,\pi$ are also shown. The life-time $\tau$ is evaluated at $T\rightarrow0$; the data for panels (a,c) and (b,d) are evaluated for $k_z=0,\pi/2$, respectively.  }
\label{fig:map}
\end{figure}

Why are the deviations in the inplane case smaller? To understand this, it is worth recalling that the band-velocities in the $x$ direction are very small at the nodes whereas they are largest there in the $z$ direction. This is shown also on the contour map on Fig.~\ref{fig:map}(a,b) that depicts $(v^\alpha_\veck)^2$ for the inplane ($\alpha=x)$ and outofplane ($\alpha=z$) cases, respectively.

That is, in the inplane case the band transport function suppresses the contribution from the nodes. There the Fermi velocitities are smaller and the condition in Eq.~\ref{eq:cond} is satisfied to a lesser degree.

One can rationalize this also from another point of view, because the transport function retains in the inplane case just momenta at the antinodes means that one can treat the scattering as momentum independent, with the magnitude given by the value at the antinode. Hence, for the inplane case one can apply the derivation of the Boltzmann transport for the case without momentum dependence~\cite{georges2021}, whereas this cannot be done for the outofplane case.
One can illustrate this qualitative distinction also by inspecting  $(v_\veck^\alpha)^2 \tau_\veck$  that is shown in (c,d) for the inplane (outofplane) case, respectively. For the inplane case panel (c) resembles panel (a) which tells the momentum dependence of scattering is not important. For the outofplane response this is not the case: the velocities are largest where the scattering is also large and the interplay between the two must be taken into account.



%


\begin{thebibliography}{43}%
\makeatletter
\providecommand \@ifxundefined [1]{%
 \@ifx{#1\undefined}
}%
\providecommand \@ifnum [1]{%
 \ifnum #1\expandafter \@firstoftwo
 \else \expandafter \@secondoftwo
 \fi
}%
\providecommand \@ifx [1]{%
 \ifx #1\expandafter \@firstoftwo
 \else \expandafter \@secondoftwo
 \fi
}%
\providecommand \natexlab [1]{#1}%
\providecommand \enquote  [1]{``#1''}%
\providecommand \bibnamefont  [1]{#1}%
\providecommand \bibfnamefont [1]{#1}%
\providecommand \citenamefont [1]{#1}%
\providecommand \href@noop [0]{\@secondoftwo}%
\providecommand \href [0]{\begingroup \@sanitize@url \@href}%
\providecommand \@href[1]{\@@startlink{#1}\@@href}%
\providecommand \@@href[1]{\endgroup#1\@@endlink}%
\providecommand \@sanitize@url [0]{\catcode `\\12\catcode `\$12\catcode
  `\&12\catcode `\#12\catcode `\^12\catcode `\_12\catcode `\%12\relax}%
\providecommand \@@startlink[1]{}%
\providecommand \@@endlink[0]{}%
\providecommand \url  [0]{\begingroup\@sanitize@url \@url }%
\providecommand \@url [1]{\endgroup\@href {#1}{\urlprefix }}%
\providecommand \urlprefix  [0]{URL }%
\providecommand \Eprint [0]{\href }%
\providecommand \doibase [0]{https://doi.org/}%
\providecommand \selectlanguage [0]{\@gobble}%
\providecommand \bibinfo  [0]{\@secondoftwo}%
\providecommand \bibfield  [0]{\@secondoftwo}%
\providecommand \translation [1]{[#1]}%
\providecommand \BibitemOpen [0]{}%
\providecommand \bibitemStop [0]{}%
\providecommand \bibitemNoStop [0]{.\EOS\space}%
\providecommand \EOS [0]{\spacefactor3000\relax}%
\providecommand \BibitemShut  [1]{\csname bibitem#1\endcsname}%
\let\auto@bib@innerbib\@empty
\bibitem [{\citenamefont {Proust}\ and\ \citenamefont
  {Taillefer}(2019)}]{proust2019}%
  \BibitemOpen
  \bibfield  {author} {\bibinfo {author} {\bibfnamefont {C.}~\bibnamefont
  {Proust}}\ and\ \bibinfo {author} {\bibfnamefont {L.}~\bibnamefont
  {Taillefer}},\ }\bibfield  {title} {\bibinfo {title} {The {Remarkable}
  {Underlying} {Ground} {States} of {Cuprate} {Superconductors}},\ }\href
  {https://doi.org/10.1146/annurev-conmatphys-031218-013210} {\bibfield
  {journal} {\bibinfo  {journal} {Annual Review of Condensed Matter Physics}\
  }\textbf {\bibinfo {volume} {10}},\ \bibinfo {pages} {409} (\bibinfo {year}
  {2019})}\BibitemShut {NoStop}%
\bibitem [{\citenamefont {Badoux}\ \emph {et~al.}(2016)\citenamefont {Badoux},
  \citenamefont {Tabis}, \citenamefont {Lalibert{\'e}}, \citenamefont
  {Grissonnanche}, \citenamefont {Vignolle}, \citenamefont {Vignolles},
  \citenamefont {B{\'e}ard}, \citenamefont {Bonn}, \citenamefont {Hardy},
  \citenamefont {Liang}, \citenamefont {{N. Doiron-Leyraud}}, \citenamefont
  {Taillefer},\ and\ \citenamefont {Proust}}]{badoux2016}%
  \BibitemOpen
  \bibfield  {author} {\bibinfo {author} {\bibfnamefont {S.}~\bibnamefont
  {Badoux}}, \bibinfo {author} {\bibfnamefont {W.}~\bibnamefont {Tabis}},
  \bibinfo {author} {\bibfnamefont {F.}~\bibnamefont {Lalibert{\'e}}}, \bibinfo
  {author} {\bibfnamefont {G.}~\bibnamefont {Grissonnanche}}, \bibinfo {author}
  {\bibfnamefont {B.}~\bibnamefont {Vignolle}}, \bibinfo {author}
  {\bibfnamefont {D.}~\bibnamefont {Vignolles}}, \bibinfo {author}
  {\bibfnamefont {J.}~\bibnamefont {B{\'e}ard}}, \bibinfo {author}
  {\bibfnamefont {D.~A.}\ \bibnamefont {Bonn}}, \bibinfo {author}
  {\bibfnamefont {W.~N.}\ \bibnamefont {Hardy}}, \bibinfo {author}
  {\bibfnamefont {R.}~\bibnamefont {Liang}}, \bibinfo {author} {\bibnamefont
  {{N. Doiron-Leyraud}}}, \bibinfo {author} {\bibfnamefont {L.}~\bibnamefont
  {Taillefer}},\ and\ \bibinfo {author} {\bibfnamefont {C.}~\bibnamefont
  {Proust}},\ }\bibfield  {title} {\bibinfo {title} {Change of carrier density
  at the pseudogap critical point of a cuprate superconductor},\ }\href
  {https://doi.org/10.1038/nature16983} {\bibfield  {journal} {\bibinfo
  {journal} {Nature}\ }\textbf {\bibinfo {volume} {531}},\ \bibinfo {pages}
  {210} (\bibinfo {year} {2016})}\BibitemShut {NoStop}%
\bibitem [{\citenamefont {Collignon}\ \emph {et~al.}(2017)\citenamefont
  {Collignon}, \citenamefont {Badoux}, \citenamefont {Afshar}, \citenamefont
  {Michon}, \citenamefont {Lalibert{\'e}}, \citenamefont {Cyr-Choini{\`e}re},
  \citenamefont {Zhou}, \citenamefont {Licciardello}, \citenamefont {Wiedmann},
  \citenamefont {Doiron-Leyraud},\ and\ \citenamefont
  {Taillefer}}]{collignon2017}%
  \BibitemOpen
  \bibfield  {author} {\bibinfo {author} {\bibfnamefont {C.}~\bibnamefont
  {Collignon}}, \bibinfo {author} {\bibfnamefont {S.}~\bibnamefont {Badoux}},
  \bibinfo {author} {\bibfnamefont {S.~A.~A.}\ \bibnamefont {Afshar}}, \bibinfo
  {author} {\bibfnamefont {B.}~\bibnamefont {Michon}}, \bibinfo {author}
  {\bibfnamefont {F.}~\bibnamefont {Lalibert{\'e}}}, \bibinfo {author}
  {\bibfnamefont {O.}~\bibnamefont {Cyr-Choini{\`e}re}}, \bibinfo {author}
  {\bibfnamefont {J.-S.}\ \bibnamefont {Zhou}}, \bibinfo {author}
  {\bibfnamefont {S.}~\bibnamefont {Licciardello}}, \bibinfo {author}
  {\bibfnamefont {S.}~\bibnamefont {Wiedmann}}, \bibinfo {author}
  {\bibfnamefont {N.}~\bibnamefont {Doiron-Leyraud}},\ and\ \bibinfo {author}
  {\bibfnamefont {L.}~\bibnamefont {Taillefer}},\ }\bibfield  {title} {\bibinfo
  {title} {Fermi-surface transformation across the pseudogap critical point of
  the cuprate superconductor {La}$_{1.6-x}${Nd}$_{0.4}${Sr}$_x${CuO}$_4$},\
  }\href {https://doi.org/10.1103/PhysRevB.95.224517} {\bibfield  {journal}
  {\bibinfo  {journal} {Physical Review B}\ }\textbf {\bibinfo {volume} {95}},\
  \bibinfo {pages} {224517} (\bibinfo {year} {2017})}\BibitemShut {NoStop}%
\bibitem [{\citenamefont {Lizaire}\ \emph {et~al.}(2021)\citenamefont
  {Lizaire}, \citenamefont {Legros}, \citenamefont {Gourgout}, \citenamefont
  {Benhabib}, \citenamefont {Badoux}, \citenamefont {Lalibert\'e},
  \citenamefont {Boulanger}, \citenamefont {Ataei}, \citenamefont
  {Grissonnanche}, \citenamefont {LeBoeuf}, \citenamefont {Licciardello},
  \citenamefont {Wiedmann}, \citenamefont {Ono}, \citenamefont {Raffy},
  \citenamefont {Kawasaki}, \citenamefont {Zheng}, \citenamefont
  {Doiron-Leyraud}, \citenamefont {Proust},\ and\ \citenamefont
  {Taillefer}}]{lizaire2020transport}%
  \BibitemOpen
  \bibfield  {author} {\bibinfo {author} {\bibfnamefont {M.}~\bibnamefont
  {Lizaire}}, \bibinfo {author} {\bibfnamefont {A.}~\bibnamefont {Legros}},
  \bibinfo {author} {\bibfnamefont {A.}~\bibnamefont {Gourgout}}, \bibinfo
  {author} {\bibfnamefont {S.}~\bibnamefont {Benhabib}}, \bibinfo {author}
  {\bibfnamefont {S.}~\bibnamefont {Badoux}}, \bibinfo {author} {\bibfnamefont
  {F.}~\bibnamefont {Lalibert\'e}}, \bibinfo {author} {\bibfnamefont {M.-E.}\
  \bibnamefont {Boulanger}}, \bibinfo {author} {\bibfnamefont {A.}~\bibnamefont
  {Ataei}}, \bibinfo {author} {\bibfnamefont {G.}~\bibnamefont
  {Grissonnanche}}, \bibinfo {author} {\bibfnamefont {D.}~\bibnamefont
  {LeBoeuf}}, \bibinfo {author} {\bibfnamefont {S.}~\bibnamefont
  {Licciardello}}, \bibinfo {author} {\bibfnamefont {S.}~\bibnamefont
  {Wiedmann}}, \bibinfo {author} {\bibfnamefont {S.}~\bibnamefont {Ono}},
  \bibinfo {author} {\bibfnamefont {H.}~\bibnamefont {Raffy}}, \bibinfo
  {author} {\bibfnamefont {S.}~\bibnamefont {Kawasaki}}, \bibinfo {author}
  {\bibfnamefont {G.-Q.}\ \bibnamefont {Zheng}}, \bibinfo {author}
  {\bibfnamefont {N.}~\bibnamefont {Doiron-Leyraud}}, \bibinfo {author}
  {\bibfnamefont {C.}~\bibnamefont {Proust}},\ and\ \bibinfo {author}
  {\bibfnamefont {L.}~\bibnamefont {Taillefer}},\ }\bibfield  {title} {\bibinfo
  {title} {Transport signatures of the pseudogap critical point in the cuprate
  superconductor {Bi}$_2${Sr}$_{2-x}${La}$_x${Cu}{O}$_{6+\delta}$},\ }\href
  {https://doi.org/10.1103/PhysRevB.104.014515} {\bibfield  {journal} {\bibinfo
   {journal} {Phys. Rev. B}\ }\textbf {\bibinfo {volume} {104}},\ \bibinfo
  {pages} {014515} (\bibinfo {year} {2021})}\BibitemShut {NoStop}%
\bibitem [{\citenamefont {Lalibert{\'e}}\ \emph {et~al.}(2016)\citenamefont
  {Lalibert{\'e}}, \citenamefont {Tabis}, \citenamefont {Badoux}, \citenamefont
  {Vignolle}, \citenamefont {Destraz}, \citenamefont {Momono}, \citenamefont
  {Kurosawa}, \citenamefont {Yamada}, \citenamefont {Takagi}, \citenamefont
  {Doiron-Leyraud}, \citenamefont {Proust},\ and\ \citenamefont
  {Taillefer}}]{laliberte2016}%
  \BibitemOpen
  \bibfield  {author} {\bibinfo {author} {\bibfnamefont {F.}~\bibnamefont
  {Lalibert{\'e}}}, \bibinfo {author} {\bibfnamefont {W.}~\bibnamefont
  {Tabis}}, \bibinfo {author} {\bibfnamefont {S.}~\bibnamefont {Badoux}},
  \bibinfo {author} {\bibfnamefont {B.}~\bibnamefont {Vignolle}}, \bibinfo
  {author} {\bibfnamefont {D.}~\bibnamefont {Destraz}}, \bibinfo {author}
  {\bibfnamefont {N.}~\bibnamefont {Momono}}, \bibinfo {author} {\bibfnamefont
  {T.}~\bibnamefont {Kurosawa}}, \bibinfo {author} {\bibfnamefont
  {K.}~\bibnamefont {Yamada}}, \bibinfo {author} {\bibfnamefont
  {H.}~\bibnamefont {Takagi}}, \bibinfo {author} {\bibfnamefont
  {N.}~\bibnamefont {Doiron-Leyraud}}, \bibinfo {author} {\bibfnamefont
  {C.}~\bibnamefont {Proust}},\ and\ \bibinfo {author} {\bibfnamefont
  {L.}~\bibnamefont {Taillefer}},\ }\bibfield  {title} {\bibinfo {title}
  {Origin of the metal-to-insulator crossover in cuprate superconductors},\
  }\href {http://arxiv.org/abs/1606.04491} {\bibfield  {journal} {\bibinfo
  {journal} {arXiv:1606.04491}\ } (\bibinfo {year} {2016})}\BibitemShut
  {NoStop}%
\bibitem [{\citenamefont {Michon}\ \emph {et~al.}(2018)\citenamefont {Michon},
  \citenamefont {Ataei}, \citenamefont {Bourgeois-Hope}, \citenamefont
  {Collignon}, \citenamefont {Li}, \citenamefont {Badoux}, \citenamefont
  {Gourgout}, \citenamefont {Lalibert{\'e}}, \citenamefont {Zhou},
  \citenamefont {Doiron-Leyraud},\ and\ \citenamefont
  {Taillefer}}]{michon2018}%
  \BibitemOpen
  \bibfield  {author} {\bibinfo {author} {\bibfnamefont {B.}~\bibnamefont
  {Michon}}, \bibinfo {author} {\bibfnamefont {A.}~\bibnamefont {Ataei}},
  \bibinfo {author} {\bibfnamefont {P.}~\bibnamefont {Bourgeois-Hope}},
  \bibinfo {author} {\bibfnamefont {C.}~\bibnamefont {Collignon}}, \bibinfo
  {author} {\bibfnamefont {S.}~\bibnamefont {Li}}, \bibinfo {author}
  {\bibfnamefont {S.}~\bibnamefont {Badoux}}, \bibinfo {author} {\bibfnamefont
  {A.}~\bibnamefont {Gourgout}}, \bibinfo {author} {\bibfnamefont
  {F.}~\bibnamefont {Lalibert{\'e}}}, \bibinfo {author} {\bibfnamefont {J.-S.}\
  \bibnamefont {Zhou}}, \bibinfo {author} {\bibfnamefont {N.}~\bibnamefont
  {Doiron-Leyraud}},\ and\ \bibinfo {author} {\bibfnamefont {L.}~\bibnamefont
  {Taillefer}},\ }\bibfield  {title} {\bibinfo {title} {Wiedemann-{Franz} law
  and abrupt change in conductivity across the {Pseudogap} critical point of a
  cuprate superconductor},\ }\href {https://doi.org/10.1103/PhysRevX.8.041010}
  {\bibfield  {journal} {\bibinfo  {journal} {Physical Review X}\ }\textbf
  {\bibinfo {volume} {8}},\ \bibinfo {pages} {041010} (\bibinfo {year}
  {2018})}\BibitemShut {NoStop}%
\bibitem [{\citenamefont {Fang}\ \emph {et~al.}(2020)\citenamefont {Fang},
  \citenamefont {Grissonnanche}, \citenamefont {Legros}, \citenamefont
  {Verret}, \citenamefont {Lalibert{\'e}}, \citenamefont {Collignon},
  \citenamefont {Ataei}, \citenamefont {Dion}, \citenamefont {Zhou},
  \citenamefont {Graf}, \citenamefont {Lawler}, \citenamefont {Goddard},
  \citenamefont {Taillefer},\ and\ \citenamefont {Ramshaw}}]{fang2020}%
  \BibitemOpen
  \bibfield  {author} {\bibinfo {author} {\bibfnamefont {Y.}~\bibnamefont
  {Fang}}, \bibinfo {author} {\bibfnamefont {G.}~\bibnamefont {Grissonnanche}},
  \bibinfo {author} {\bibfnamefont {A.}~\bibnamefont {Legros}}, \bibinfo
  {author} {\bibfnamefont {S.}~\bibnamefont {Verret}}, \bibinfo {author}
  {\bibfnamefont {F.}~\bibnamefont {Lalibert{\'e}}}, \bibinfo {author}
  {\bibfnamefont {C.}~\bibnamefont {Collignon}}, \bibinfo {author}
  {\bibfnamefont {A.}~\bibnamefont {Ataei}}, \bibinfo {author} {\bibfnamefont
  {M.}~\bibnamefont {Dion}}, \bibinfo {author} {\bibfnamefont {J.}~\bibnamefont
  {Zhou}}, \bibinfo {author} {\bibfnamefont {D.}~\bibnamefont {Graf}}, \bibinfo
  {author} {\bibfnamefont {M.~J.}\ \bibnamefont {Lawler}}, \bibinfo {author}
  {\bibfnamefont {P.}~\bibnamefont {Goddard}}, \bibinfo {author} {\bibfnamefont
  {L.}~\bibnamefont {Taillefer}},\ and\ \bibinfo {author} {\bibfnamefont
  {B.~J.}\ \bibnamefont {Ramshaw}},\ }\bibfield  {title} {\bibinfo {title}
  {Fermi surface transformation at the pseudogap critical point of a cuprate
  superconductor},\ }\href {http://arxiv.org/abs/2004.01725} {\bibfield
  {journal} {\bibinfo  {journal} {arXiv:2004.01725}\ } (\bibinfo {year}
  {2020})}\BibitemShut {NoStop}%
\bibitem [{\citenamefont {Michon}\ \emph {et~al.}(2019)\citenamefont {Michon},
  \citenamefont {Girod}, \citenamefont {Badoux}, \citenamefont {Ka{\v c}mar{\v
  c}{\'\i}k}, \citenamefont {Ma}, \citenamefont {Dragomir}, \citenamefont
  {Dabkowska}, \citenamefont {Gaulin}, \citenamefont {Zhou}, \citenamefont
  {Pyon}, \citenamefont {Takayama}, \citenamefont {Takagi}, \citenamefont
  {Verret}, \citenamefont {Doiron-Leyraud}, \citenamefont {Marcenat},
  \citenamefont {Taillefer},\ and\ \citenamefont {Klein}}]{michon2019}%
  \BibitemOpen
  \bibfield  {author} {\bibinfo {author} {\bibfnamefont {B.}~\bibnamefont
  {Michon}}, \bibinfo {author} {\bibfnamefont {C.}~\bibnamefont {Girod}},
  \bibinfo {author} {\bibfnamefont {S.}~\bibnamefont {Badoux}}, \bibinfo
  {author} {\bibfnamefont {J.}~\bibnamefont {Ka{\v c}mar{\v c}{\'\i}k}},
  \bibinfo {author} {\bibfnamefont {Q.}~\bibnamefont {Ma}}, \bibinfo {author}
  {\bibfnamefont {M.}~\bibnamefont {Dragomir}}, \bibinfo {author}
  {\bibfnamefont {H.~A.}\ \bibnamefont {Dabkowska}}, \bibinfo {author}
  {\bibfnamefont {B.~D.}\ \bibnamefont {Gaulin}}, \bibinfo {author}
  {\bibfnamefont {J.-S.}\ \bibnamefont {Zhou}}, \bibinfo {author}
  {\bibfnamefont {S.}~\bibnamefont {Pyon}}, \bibinfo {author} {\bibfnamefont
  {T.}~\bibnamefont {Takayama}}, \bibinfo {author} {\bibfnamefont
  {H.}~\bibnamefont {Takagi}}, \bibinfo {author} {\bibfnamefont
  {S.}~\bibnamefont {Verret}}, \bibinfo {author} {\bibfnamefont
  {N.}~\bibnamefont {Doiron-Leyraud}}, \bibinfo {author} {\bibfnamefont
  {C.}~\bibnamefont {Marcenat}}, \bibinfo {author} {\bibfnamefont
  {L.}~\bibnamefont {Taillefer}},\ and\ \bibinfo {author} {\bibfnamefont
  {T.}~\bibnamefont {Klein}},\ }\bibfield  {title} {\bibinfo {title}
  {Thermodynamic signatures of quantum criticality in cuprate
  superconductors},\ }\href {https://doi.org/10.1038/s41586-019-0932-x}
  {\bibfield  {journal} {\bibinfo  {journal} {Nature}\ }\textbf {\bibinfo
  {volume} {567}},\ \bibinfo {pages} {218} (\bibinfo {year}
  {2019})}\BibitemShut {NoStop}%
\bibitem [{\citenamefont {Girod}\ \emph {et~al.}(2021)\citenamefont {Girod},
  \citenamefont {LeBoeuf}, \citenamefont {Demuer}, \citenamefont {Seyfarth},
  \citenamefont {Imajo}, \citenamefont {Kindo}, \citenamefont {Kohama},
  \citenamefont {Lizaire}, \citenamefont {Legros}, \citenamefont {Gourgout},
  \citenamefont {Takagi}, \citenamefont {Kurosawa}, \citenamefont {Oda},
  \citenamefont {Momono}, \citenamefont {Chang}, \citenamefont {Ono},
  \citenamefont {Zheng}, \citenamefont {Marcenat}, \citenamefont {Taillefer},\
  and\ \citenamefont {Klein}}]{girod2021normal}%
  \BibitemOpen
  \bibfield  {author} {\bibinfo {author} {\bibfnamefont {C.}~\bibnamefont
  {Girod}}, \bibinfo {author} {\bibfnamefont {D.}~\bibnamefont {LeBoeuf}},
  \bibinfo {author} {\bibfnamefont {A.}~\bibnamefont {Demuer}}, \bibinfo
  {author} {\bibfnamefont {G.}~\bibnamefont {Seyfarth}}, \bibinfo {author}
  {\bibfnamefont {S.}~\bibnamefont {Imajo}}, \bibinfo {author} {\bibfnamefont
  {K.}~\bibnamefont {Kindo}}, \bibinfo {author} {\bibfnamefont
  {Y.}~\bibnamefont {Kohama}}, \bibinfo {author} {\bibfnamefont
  {M.}~\bibnamefont {Lizaire}}, \bibinfo {author} {\bibfnamefont
  {A.}~\bibnamefont {Legros}}, \bibinfo {author} {\bibfnamefont
  {A.}~\bibnamefont {Gourgout}}, \bibinfo {author} {\bibfnamefont
  {H.}~\bibnamefont {Takagi}}, \bibinfo {author} {\bibfnamefont
  {T.}~\bibnamefont {Kurosawa}}, \bibinfo {author} {\bibfnamefont
  {M.}~\bibnamefont {Oda}}, \bibinfo {author} {\bibfnamefont {N.}~\bibnamefont
  {Momono}}, \bibinfo {author} {\bibfnamefont {J.}~\bibnamefont {Chang}},
  \bibinfo {author} {\bibfnamefont {S.}~\bibnamefont {Ono}}, \bibinfo {author}
  {\bibfnamefont {G.-q.}\ \bibnamefont {Zheng}}, \bibinfo {author}
  {\bibfnamefont {C.}~\bibnamefont {Marcenat}}, \bibinfo {author}
  {\bibfnamefont {L.}~\bibnamefont {Taillefer}},\ and\ \bibinfo {author}
  {\bibfnamefont {T.}~\bibnamefont {Klein}},\ }\bibfield  {title} {\bibinfo
  {title} {Normal state specific heat in the cuprate superconductors
  {La}$_{2-x}${Sr}$_x${CuO}$_4$ and
  {Bi}$_{2+y}${Sr}$_{2-x-y}${La}$_x${CuO}$_{6+\delta}$ near the critical point
  of the pseudogap phase},\ }\href
  {https://doi.org/10.1103/PhysRevB.103.214506} {\bibfield  {journal} {\bibinfo
   {journal} {Phys. Rev. B}\ }\textbf {\bibinfo {volume} {103}},\ \bibinfo
  {pages} {214506} (\bibinfo {year} {2021})}\BibitemShut {NoStop}%
\bibitem [{\citenamefont {Collignon}\ \emph {et~al.}(2021)\citenamefont
  {Collignon}, \citenamefont {Ataei}, \citenamefont {Gourgout}, \citenamefont
  {Badoux}, \citenamefont {Lizaire}, \citenamefont {Legros}, \citenamefont
  {Licciardello}, \citenamefont {Wiedmann}, \citenamefont {Yan}, \citenamefont
  {Zhou}, \citenamefont {Ma}, \citenamefont {Gaulin}, \citenamefont
  {Doiron-Leyraud},\ and\ \citenamefont
  {Taillefer}}]{collignon2021thermopower}%
  \BibitemOpen
  \bibfield  {author} {\bibinfo {author} {\bibfnamefont {C.}~\bibnamefont
  {Collignon}}, \bibinfo {author} {\bibfnamefont {A.}~\bibnamefont {Ataei}},
  \bibinfo {author} {\bibfnamefont {A.}~\bibnamefont {Gourgout}}, \bibinfo
  {author} {\bibfnamefont {S.}~\bibnamefont {Badoux}}, \bibinfo {author}
  {\bibfnamefont {M.}~\bibnamefont {Lizaire}}, \bibinfo {author} {\bibfnamefont
  {A.}~\bibnamefont {Legros}}, \bibinfo {author} {\bibfnamefont
  {S.}~\bibnamefont {Licciardello}}, \bibinfo {author} {\bibfnamefont
  {S.}~\bibnamefont {Wiedmann}}, \bibinfo {author} {\bibfnamefont {J.-Q.}\
  \bibnamefont {Yan}}, \bibinfo {author} {\bibfnamefont {J.-S.}\ \bibnamefont
  {Zhou}}, \bibinfo {author} {\bibfnamefont {Q.}~\bibnamefont {Ma}}, \bibinfo
  {author} {\bibfnamefont {B.~D.}\ \bibnamefont {Gaulin}}, \bibinfo {author}
  {\bibfnamefont {N.}~\bibnamefont {Doiron-Leyraud}},\ and\ \bibinfo {author}
  {\bibfnamefont {L.}~\bibnamefont {Taillefer}},\ }\bibfield  {title} {\bibinfo
  {title} {Thermopower across the phase diagram of the cuprate
  {La}$_{1.6-x}${Nd}$_{0.4}${Sr}$_x${CuO}$_4$ : signatures of the pseudogap and
  charge-density-wave phases},\ }\href
  {https://doi.org/10.1103/PhysRevB.103.155102} {\bibfield  {journal} {\bibinfo
   {journal} {Phys. Rev. B}\ }\textbf {\bibinfo {volume} {103}},\ \bibinfo
  {pages} {155102} (\bibinfo {year} {2021})}\BibitemShut {NoStop}%
\bibitem [{\citenamefont {Behnia}\ \emph {et~al.}(2004)\citenamefont {Behnia},
  \citenamefont {Jaccard},\ and\ \citenamefont {Flouquet}}]{behnia2004}%
  \BibitemOpen
  \bibfield  {author} {\bibinfo {author} {\bibfnamefont {K.}~\bibnamefont
  {Behnia}}, \bibinfo {author} {\bibfnamefont {D.}~\bibnamefont {Jaccard}},\
  and\ \bibinfo {author} {\bibfnamefont {J.}~\bibnamefont {Flouquet}},\
  }\bibfield  {title} {\bibinfo {title} {On the thermoelectricity of correlated
  electrons in the zero-temperature limit},\ }\href
  {http://iopscience.iop.org/0953-8984/16/28/037} {\bibfield  {journal}
  {\bibinfo  {journal} {Journal of Physics: Condensed Matter}\ }\textbf
  {\bibinfo {volume} {16}},\ \bibinfo {pages} {5187} (\bibinfo {year}
  {2004})}\BibitemShut {NoStop}%
\bibitem [{\citenamefont {Verret}\ \emph {et~al.}(2017)\citenamefont {Verret},
  \citenamefont {Simard}, \citenamefont {Charlebois}, \citenamefont
  {S{\'e}n{\'e}chal},\ and\ \citenamefont {Tremblay}}]{verret2017}%
  \BibitemOpen
  \bibfield  {author} {\bibinfo {author} {\bibfnamefont {S.}~\bibnamefont
  {Verret}}, \bibinfo {author} {\bibfnamefont {O.}~\bibnamefont {Simard}},
  \bibinfo {author} {\bibfnamefont {M.}~\bibnamefont {Charlebois}}, \bibinfo
  {author} {\bibfnamefont {D.}~\bibnamefont {S{\'e}n{\'e}chal}},\ and\ \bibinfo
  {author} {\bibfnamefont {A.-M.~S.}\ \bibnamefont {Tremblay}},\ }\bibfield
  {title} {\bibinfo {title} {Phenomenological theories of the low-temperature
  pseudogap: {Hall} number, specific heat, and {Seebeck} coefficient},\ }\href
  {https://doi.org/10.1103/PhysRevB.96.125139} {\bibfield  {journal} {\bibinfo
  {journal} {Physical Review B}\ }\textbf {\bibinfo {volume} {96}},\ \bibinfo
  {pages} {125139} (\bibinfo {year} {2017})}\BibitemShut {NoStop}%
\bibitem [{\citenamefont {Behnia}(2015)}]{behnia_2015}%
  \BibitemOpen
  \bibfield  {author} {\bibinfo {author} {\bibfnamefont {K.}~\bibnamefont
  {Behnia}},\ }\href@noop {} {\bibfield  {journal} {\bibinfo  {journal}
  {Fundamentals of thermoelectricity}\ } (\bibinfo {year} {2015})}\BibitemShut
  {NoStop}%
\bibitem [{\citenamefont {Kondo}\ \emph {et~al.}(2005)\citenamefont {Kondo},
  \citenamefont {Takeuchi}, \citenamefont {Mizutani}, \citenamefont {Yokoya},
  \citenamefont {Tsuda},\ and\ \citenamefont {Shin}}]{kondo2005}%
  \BibitemOpen
  \bibfield  {author} {\bibinfo {author} {\bibfnamefont {T.}~\bibnamefont
  {Kondo}}, \bibinfo {author} {\bibfnamefont {T.}~\bibnamefont {Takeuchi}},
  \bibinfo {author} {\bibfnamefont {U.}~\bibnamefont {Mizutani}}, \bibinfo
  {author} {\bibfnamefont {T.}~\bibnamefont {Yokoya}}, \bibinfo {author}
  {\bibfnamefont {S.}~\bibnamefont {Tsuda}},\ and\ \bibinfo {author}
  {\bibfnamefont {S.}~\bibnamefont {Shin}},\ }\bibfield  {title} {\bibinfo
  {title} {Contribution of electronic structure to thermoelectric power in
  {Bi}{Pb}$_2${Sr}{La}$_2${Cu}{O}$_{6+\delta}$},\ }\href
  {https://doi.org/10.1103/PhysRevB.72.024533} {\bibfield  {journal} {\bibinfo
  {journal} {Physical Review B}\ }\textbf {\bibinfo {volume} {72}},\ \bibinfo
  {pages} {024533} (\bibinfo {year} {2005})}\BibitemShut {NoStop}%
\bibitem [{\citenamefont {Matt}\ \emph {et~al.}(2015)\citenamefont {Matt},
  \citenamefont {Fatuzzo}, \citenamefont {Sassa}, \citenamefont {M{\aa}nsson},
  \citenamefont {Fatale}, \citenamefont {Bitetta}, \citenamefont {Shi},
  \citenamefont {Pailh{\`e}s}, \citenamefont {Berntsen}, \citenamefont
  {Kurosawa}, \citenamefont {Oda}, \citenamefont {Momono}, \citenamefont
  {Lipscombe}, \citenamefont {Hayden}, \citenamefont {Yan}, \citenamefont
  {Zhou}, \citenamefont {Goodenough}, \citenamefont {Pyon}, \citenamefont
  {Takayama}, \citenamefont {Takagi}, \citenamefont {Patthey}, \citenamefont
  {Bendounan}, \citenamefont {Razzoli}, \citenamefont {Shi}, \citenamefont
  {Plumb}, \citenamefont {Radovic}, \citenamefont {Grioni}, \citenamefont
  {Mesot}, \citenamefont {Tjernberg},\ and\ \citenamefont {Chang}}]{matt2015}%
  \BibitemOpen
  \bibfield  {author} {\bibinfo {author} {\bibfnamefont {C.~E.}\ \bibnamefont
  {Matt}}, \bibinfo {author} {\bibfnamefont {C.~G.}\ \bibnamefont {Fatuzzo}},
  \bibinfo {author} {\bibfnamefont {Y.}~\bibnamefont {Sassa}}, \bibinfo
  {author} {\bibfnamefont {M.}~\bibnamefont {M{\aa}nsson}}, \bibinfo {author}
  {\bibfnamefont {S.}~\bibnamefont {Fatale}}, \bibinfo {author} {\bibfnamefont
  {V.}~\bibnamefont {Bitetta}}, \bibinfo {author} {\bibfnamefont
  {X.}~\bibnamefont {Shi}}, \bibinfo {author} {\bibfnamefont {S.}~\bibnamefont
  {Pailh{\`e}s}}, \bibinfo {author} {\bibfnamefont {M.~H.}\ \bibnamefont
  {Berntsen}}, \bibinfo {author} {\bibfnamefont {T.}~\bibnamefont {Kurosawa}},
  \bibinfo {author} {\bibfnamefont {M.}~\bibnamefont {Oda}}, \bibinfo {author}
  {\bibfnamefont {N.}~\bibnamefont {Momono}}, \bibinfo {author} {\bibfnamefont
  {O.~J.}\ \bibnamefont {Lipscombe}}, \bibinfo {author} {\bibfnamefont {S.~M.}\
  \bibnamefont {Hayden}}, \bibinfo {author} {\bibfnamefont {J.-Q.}\
  \bibnamefont {Yan}}, \bibinfo {author} {\bibfnamefont {J.-S.}\ \bibnamefont
  {Zhou}}, \bibinfo {author} {\bibfnamefont {J.~B.}\ \bibnamefont
  {Goodenough}}, \bibinfo {author} {\bibfnamefont {S.}~\bibnamefont {Pyon}},
  \bibinfo {author} {\bibfnamefont {T.}~\bibnamefont {Takayama}}, \bibinfo
  {author} {\bibfnamefont {H.}~\bibnamefont {Takagi}}, \bibinfo {author}
  {\bibfnamefont {L.}~\bibnamefont {Patthey}}, \bibinfo {author} {\bibfnamefont
  {A.}~\bibnamefont {Bendounan}}, \bibinfo {author} {\bibfnamefont
  {E.}~\bibnamefont {Razzoli}}, \bibinfo {author} {\bibfnamefont
  {M.}~\bibnamefont {Shi}}, \bibinfo {author} {\bibfnamefont {N.~C.}\
  \bibnamefont {Plumb}}, \bibinfo {author} {\bibfnamefont {M.}~\bibnamefont
  {Radovic}}, \bibinfo {author} {\bibfnamefont {M.}~\bibnamefont {Grioni}},
  \bibinfo {author} {\bibfnamefont {J.}~\bibnamefont {Mesot}}, \bibinfo
  {author} {\bibfnamefont {O.}~\bibnamefont {Tjernberg}},\ and\ \bibinfo
  {author} {\bibfnamefont {J.}~\bibnamefont {Chang}},\ }\bibfield  {title}
  {\bibinfo {title} {Electron scattering, charge order, and pseudogap physics
  in {La}$_{1.6-x}${Nd}$_{0.4}${Sr}$_x${CuO}$_4$: {An} angle-resolved
  photoemission spectroscopy study},\ }\href
  {https://doi.org/10.1103/PhysRevB.92.134524} {\bibfield  {journal} {\bibinfo
  {journal} {Physical Review B}\ }\textbf {\bibinfo {volume} {92}},\ \bibinfo
  {pages} {134524} (\bibinfo {year} {2015})}\BibitemShut {NoStop}%
\bibitem [{\citenamefont {Grissonnanche}\ \emph {et~al.}(2021)\citenamefont
  {Grissonnanche}, \citenamefont {Fang}, \citenamefont {Legros}, \citenamefont
  {Verret}, \citenamefont {Laliberté}, \citenamefont {Collignon},
  \citenamefont {Zhou}, \citenamefont {Graf}, \citenamefont {Goddard},
  \citenamefont {Taillefer},\ and\ \citenamefont
  {Ramshaw}}]{grissonnanche2020b}%
  \BibitemOpen
  \bibfield  {author} {\bibinfo {author} {\bibfnamefont {G.}~\bibnamefont
  {Grissonnanche}}, \bibinfo {author} {\bibfnamefont {Y.}~\bibnamefont {Fang}},
  \bibinfo {author} {\bibfnamefont {A.}~\bibnamefont {Legros}}, \bibinfo
  {author} {\bibfnamefont {S.}~\bibnamefont {Verret}}, \bibinfo {author}
  {\bibfnamefont {F.}~\bibnamefont {Laliberté}}, \bibinfo {author}
  {\bibfnamefont {C.}~\bibnamefont {Collignon}}, \bibinfo {author}
  {\bibfnamefont {J.}~\bibnamefont {Zhou}}, \bibinfo {author} {\bibfnamefont
  {D.}~\bibnamefont {Graf}}, \bibinfo {author} {\bibfnamefont {P.~A.}\
  \bibnamefont {Goddard}}, \bibinfo {author} {\bibfnamefont {L.}~\bibnamefont
  {Taillefer}},\ and\ \bibinfo {author} {\bibfnamefont {B.~J.}\ \bibnamefont
  {Ramshaw}},\ }\bibfield  {title} {\bibinfo {title} {Linear-in temperature
  resistivity from an isotropic {Planckian} scattering rate},\ }\href
  {https://doi.org/10.1038/s41586-021-03697-8} {\bibfield  {journal} {\bibinfo
  {journal} {Nature}\ }\textbf {\bibinfo {volume} {595}},\ \bibinfo {pages}
  {667} (\bibinfo {year} {2021})}\BibitemShut {NoStop}%
\bibitem [{\citenamefont {Daou}\ \emph
  {et~al.}(2009{\natexlab{a}})\citenamefont {Daou}, \citenamefont
  {Doiron-Leyraud}, \citenamefont {LeBoeuf}, \citenamefont {Li}, \citenamefont
  {Lalibert{\'e}}, \citenamefont {Cyr-Choini{\`e}re}, \citenamefont {Jo},
  \citenamefont {Balicas}, \citenamefont {Yan}, \citenamefont {Zhou},
  \citenamefont {Goodenough},\ and\ \citenamefont {Taillefer}}]{daou2009}%
  \BibitemOpen
  \bibfield  {author} {\bibinfo {author} {\bibfnamefont {R.}~\bibnamefont
  {Daou}}, \bibinfo {author} {\bibfnamefont {N.}~\bibnamefont
  {Doiron-Leyraud}}, \bibinfo {author} {\bibfnamefont {D.}~\bibnamefont
  {LeBoeuf}}, \bibinfo {author} {\bibfnamefont {S.~Y.}\ \bibnamefont {Li}},
  \bibinfo {author} {\bibfnamefont {F.}~\bibnamefont {Lalibert{\'e}}}, \bibinfo
  {author} {\bibfnamefont {O.}~\bibnamefont {Cyr-Choini{\`e}re}}, \bibinfo
  {author} {\bibfnamefont {Y.~J.}\ \bibnamefont {Jo}}, \bibinfo {author}
  {\bibfnamefont {L.}~\bibnamefont {Balicas}}, \bibinfo {author} {\bibfnamefont
  {J.-Q.}\ \bibnamefont {Yan}}, \bibinfo {author} {\bibfnamefont {J.-S.}\
  \bibnamefont {Zhou}}, \bibinfo {author} {\bibfnamefont {J.~B.}\ \bibnamefont
  {Goodenough}},\ and\ \bibinfo {author} {\bibfnamefont {L.}~\bibnamefont
  {Taillefer}},\ }\bibfield  {title} {\bibinfo {title} {Linear temperature
  dependence of resistivity and change in the {Fermi} surface at the pseudogap
  critical point of a high-${T}_c$ superconductor},\ }\href
  {http://www.nature.com/doifinder/10.1038/nphys1109} {\bibfield  {journal}
  {\bibinfo  {journal} {Nature Physics}\ }\textbf {\bibinfo {volume} {5}},\
  \bibinfo {pages} {31} (\bibinfo {year} {2009}{\natexlab{a}})}\BibitemShut
  {NoStop}%
\bibitem [{\citenamefont {Wang}\ \emph {et~al.}(2019)\citenamefont {Wang},
  \citenamefont {Yang}, \citenamefont {Guo}, \citenamefont {Peng},
  \citenamefont {Wang}, \citenamefont {Chu},\ and\ \citenamefont
  {Zheng}}]{wang2019}%
  \BibitemOpen
  \bibfield  {author} {\bibinfo {author} {\bibfnamefont {H.}~\bibnamefont
  {Wang}}, \bibinfo {author} {\bibfnamefont {F.}~\bibnamefont {Yang}}, \bibinfo
  {author} {\bibfnamefont {Y.}~\bibnamefont {Guo}}, \bibinfo {author}
  {\bibfnamefont {K.}~\bibnamefont {Peng}}, \bibinfo {author} {\bibfnamefont
  {D.}~\bibnamefont {Wang}}, \bibinfo {author} {\bibfnamefont {W.}~\bibnamefont
  {Chu}},\ and\ \bibinfo {author} {\bibfnamefont {S.}~\bibnamefont {Zheng}},\
  }\bibfield  {title} {\bibinfo {title} {Determination of the thermopower of
  microscale samples with an {A}{C} method},\ }\href
  {https://linkinghub.elsevier.com/retrieve/pii/S0263224118307735} {\bibfield
  {journal} {\bibinfo  {journal} {Measurement}\ }\textbf {\bibinfo {volume}
  {131}},\ \bibinfo {pages} {204} (\bibinfo {year} {2019})}\BibitemShut
  {NoStop}%
\bibitem [{\citenamefont {Daou}\ \emph
  {et~al.}(2009{\natexlab{b}})\citenamefont {Daou}, \citenamefont
  {Cyr-Choini{\`e}re}, \citenamefont {Lalibert{\'e}}, \citenamefont {LeBoeuf},
  \citenamefont {Doiron-Leyraud}, \citenamefont {Yan}, \citenamefont {Zhou},
  \citenamefont {Goodenough},\ and\ \citenamefont {Taillefer}}]{daou2009S}%
  \BibitemOpen
  \bibfield  {author} {\bibinfo {author} {\bibfnamefont {R.}~\bibnamefont
  {Daou}}, \bibinfo {author} {\bibfnamefont {O.}~\bibnamefont
  {Cyr-Choini{\`e}re}}, \bibinfo {author} {\bibfnamefont {F.}~\bibnamefont
  {Lalibert{\'e}}}, \bibinfo {author} {\bibfnamefont {D.}~\bibnamefont
  {LeBoeuf}}, \bibinfo {author} {\bibfnamefont {N.}~\bibnamefont
  {Doiron-Leyraud}}, \bibinfo {author} {\bibfnamefont {J.-Q.}\ \bibnamefont
  {Yan}}, \bibinfo {author} {\bibfnamefont {J.-S.}\ \bibnamefont {Zhou}},
  \bibinfo {author} {\bibfnamefont {J.}~\bibnamefont {Goodenough}},\ and\
  \bibinfo {author} {\bibfnamefont {L.}~\bibnamefont {Taillefer}},\ }\bibfield
  {title} {\bibinfo {title} {Thermopower across the stripe critical point of
  {La}$_{1.6-x}${Nd}$_{0.4}${Sr}$_x${CuO}$_4$: Evidence for a quantum critical
  point in a hole-doped high-${T}_c$ superconductor},\ }\href
  {http://link.aps.org/doi/10.1103/PhysRevB.79.180505} {\bibfield  {journal}
  {\bibinfo  {journal} {Physical Review B}\ }\textbf {\bibinfo {volume} {79}},\
  \bibinfo {pages} {180505(R)} (\bibinfo {year}
  {2009}{\natexlab{b}})}\BibitemShut {NoStop}%
\bibitem [{\citenamefont {Lalibert{\'e}}\ \emph {et~al.}(2011)\citenamefont
  {Lalibert{\'e}}, \citenamefont {Chang}, \citenamefont {Doiron-Leyraud},
  \citenamefont {Hassinger}, \citenamefont {Daou}, \citenamefont {Rondeau},
  \citenamefont {Ramshaw}, \citenamefont {Liang}, \citenamefont {Bonn},
  \citenamefont {Hardy}, \citenamefont {Pyon}, \citenamefont {Takayama},
  \citenamefont {Takagi}, \citenamefont {Sheikin}, \citenamefont {Malone},
  \citenamefont {Proust}, \citenamefont {Behnia},\ and\ \citenamefont
  {Taillefer}}]{laliberte2011}%
  \BibitemOpen
  \bibfield  {author} {\bibinfo {author} {\bibfnamefont {F.}~\bibnamefont
  {Lalibert{\'e}}}, \bibinfo {author} {\bibfnamefont {J.}~\bibnamefont
  {Chang}}, \bibinfo {author} {\bibfnamefont {N.}~\bibnamefont
  {Doiron-Leyraud}}, \bibinfo {author} {\bibfnamefont {E.}~\bibnamefont
  {Hassinger}}, \bibinfo {author} {\bibfnamefont {R.}~\bibnamefont {Daou}},
  \bibinfo {author} {\bibfnamefont {M.}~\bibnamefont {Rondeau}}, \bibinfo
  {author} {\bibfnamefont {B.}~\bibnamefont {Ramshaw}}, \bibinfo {author}
  {\bibfnamefont {R.}~\bibnamefont {Liang}}, \bibinfo {author} {\bibfnamefont
  {D.}~\bibnamefont {Bonn}}, \bibinfo {author} {\bibfnamefont {W.}~\bibnamefont
  {Hardy}}, \bibinfo {author} {\bibfnamefont {S.}~\bibnamefont {Pyon}},
  \bibinfo {author} {\bibfnamefont {T.}~\bibnamefont {Takayama}}, \bibinfo
  {author} {\bibfnamefont {H.}~\bibnamefont {Takagi}}, \bibinfo {author}
  {\bibfnamefont {I.}~\bibnamefont {Sheikin}}, \bibinfo {author} {\bibfnamefont
  {L.}~\bibnamefont {Malone}}, \bibinfo {author} {\bibfnamefont
  {C.}~\bibnamefont {Proust}}, \bibinfo {author} {\bibfnamefont
  {K.}~\bibnamefont {Behnia}},\ and\ \bibinfo {author} {\bibfnamefont
  {L.}~\bibnamefont {Taillefer}},\ }\bibfield  {title} {\bibinfo {title}
  {Fermi-surface reconstruction by stripe order in cuprate superconductors},\
  }\href {https://doi.org/10.1038/ncomms1440} {\bibfield  {journal} {\bibinfo
  {journal} {Nature Communications}\ }\textbf {\bibinfo {volume} {2}},\
  \bibinfo {pages} {432} (\bibinfo {year} {2011})}\BibitemShut {NoStop}%
\bibitem [{\citenamefont {Grissonnanche}\ \emph {et~al.}(2019)\citenamefont
  {Grissonnanche}, \citenamefont {Legros}, \citenamefont {Badoux},
  \citenamefont {Lefran{\c c}ois}, \citenamefont {Zatko}, \citenamefont
  {Lizaire}, \citenamefont {Lalibert{\'e}}, \citenamefont {Gourgout},
  \citenamefont {Zhou}, \citenamefont {Pyon}, \citenamefont {Takayama},
  \citenamefont {Takagi}, \citenamefont {Ono}, \citenamefont {Doiron-Leyraud},\
  and\ \citenamefont {Taillefer}}]{grissonnanche2019}%
  \BibitemOpen
  \bibfield  {author} {\bibinfo {author} {\bibfnamefont {G.}~\bibnamefont
  {Grissonnanche}}, \bibinfo {author} {\bibfnamefont {A.}~\bibnamefont
  {Legros}}, \bibinfo {author} {\bibfnamefont {S.}~\bibnamefont {Badoux}},
  \bibinfo {author} {\bibfnamefont {E.}~\bibnamefont {Lefran{\c c}ois}},
  \bibinfo {author} {\bibfnamefont {V.}~\bibnamefont {Zatko}}, \bibinfo
  {author} {\bibfnamefont {M.}~\bibnamefont {Lizaire}}, \bibinfo {author}
  {\bibfnamefont {F.}~\bibnamefont {Lalibert{\'e}}}, \bibinfo {author}
  {\bibfnamefont {A.}~\bibnamefont {Gourgout}}, \bibinfo {author}
  {\bibfnamefont {J.-S.}\ \bibnamefont {Zhou}}, \bibinfo {author}
  {\bibfnamefont {S.}~\bibnamefont {Pyon}}, \bibinfo {author} {\bibfnamefont
  {T.}~\bibnamefont {Takayama}}, \bibinfo {author} {\bibfnamefont
  {H.}~\bibnamefont {Takagi}}, \bibinfo {author} {\bibfnamefont
  {S.}~\bibnamefont {Ono}}, \bibinfo {author} {\bibfnamefont {N.}~\bibnamefont
  {Doiron-Leyraud}},\ and\ \bibinfo {author} {\bibfnamefont {L.}~\bibnamefont
  {Taillefer}},\ }\bibfield  {title} {\bibinfo {title} {Giant thermal {Hall}
  conductivity in the pseudogap phase of cuprate superconductors},\ }\href
  {https://doi.org/10.1038/s41586-019-1375-0} {\bibfield  {journal} {\bibinfo
  {journal} {Nature}\ }\textbf {\bibinfo {volume} {571}},\ \bibinfo {pages}
  {376} (\bibinfo {year} {2019})}\BibitemShut {NoStop}%
\bibitem [{\citenamefont {Grissonnanche}\ \emph {et~al.}(2020)\citenamefont
  {Grissonnanche}, \citenamefont {Thériault}, \citenamefont {Gourgout},
  \citenamefont {Boulanger}, \citenamefont {Lefrançois}, \citenamefont
  {Ataei}, \citenamefont {Laliberté}, \citenamefont {Dion}, \citenamefont
  {Zhou}, \citenamefont {Pyon}, \citenamefont {Takayama}, \citenamefont
  {Takagi}, \citenamefont {Doiron-Leyraud},\ and\ \citenamefont
  {Taillefer}}]{grissonnanche2020}%
  \BibitemOpen
  \bibfield  {author} {\bibinfo {author} {\bibfnamefont {G.}~\bibnamefont
  {Grissonnanche}}, \bibinfo {author} {\bibfnamefont {S.}~\bibnamefont
  {Thériault}}, \bibinfo {author} {\bibfnamefont {A.}~\bibnamefont
  {Gourgout}}, \bibinfo {author} {\bibfnamefont {M.-E.}\ \bibnamefont
  {Boulanger}}, \bibinfo {author} {\bibfnamefont {E.}~\bibnamefont
  {Lefrançois}}, \bibinfo {author} {\bibfnamefont {A.}~\bibnamefont {Ataei}},
  \bibinfo {author} {\bibfnamefont {F.}~\bibnamefont {Laliberté}}, \bibinfo
  {author} {\bibfnamefont {M.}~\bibnamefont {Dion}}, \bibinfo {author}
  {\bibfnamefont {J.-S.}\ \bibnamefont {Zhou}}, \bibinfo {author}
  {\bibfnamefont {S.}~\bibnamefont {Pyon}}, \bibinfo {author} {\bibfnamefont
  {T.}~\bibnamefont {Takayama}}, \bibinfo {author} {\bibfnamefont
  {H.}~\bibnamefont {Takagi}}, \bibinfo {author} {\bibfnamefont
  {N.}~\bibnamefont {Doiron-Leyraud}},\ and\ \bibinfo {author} {\bibfnamefont
  {L.}~\bibnamefont {Taillefer}},\ }\bibfield  {title} {\bibinfo {title}
  {Chiral phonons in the pseudogap phase of cuprates},\ }\href
  {https://doi.org/10.1038/s41567-020-0965-y} {\bibfield  {journal} {\bibinfo
  {journal} {Nature Physics}\ }\textbf {\bibinfo {volume} {16}},\ \bibinfo
  {pages} {1108} (\bibinfo {year} {2020})}\BibitemShut {NoStop}%
\bibitem [{\citenamefont {Obertelli}\ \emph {et~al.}(1992)\citenamefont
  {Obertelli}, \citenamefont {Cooper},\ and\ \citenamefont
  {Tallon}}]{obertelli1992}%
  \BibitemOpen
  \bibfield  {author} {\bibinfo {author} {\bibfnamefont {S.~D.}\ \bibnamefont
  {Obertelli}}, \bibinfo {author} {\bibfnamefont {J.~R.}\ \bibnamefont
  {Cooper}},\ and\ \bibinfo {author} {\bibfnamefont {J.~L.}\ \bibnamefont
  {Tallon}},\ }\bibfield  {title} {\bibinfo {title} {Systematics in the
  thermoelectric power of high-${T}_{c}$ oxides},\ }\href
  {https://doi.org/10.1103/PhysRevB.46.14928} {\bibfield  {journal} {\bibinfo
  {journal} {Phys. Rev. B}\ }\textbf {\bibinfo {volume} {46}},\ \bibinfo
  {pages} {14928} (\bibinfo {year} {1992})}\BibitemShut {NoStop}%
\bibitem [{\citenamefont {Munakata}\ \emph {et~al.}(1992)\citenamefont
  {Munakata}, \citenamefont {Matsuura}, \citenamefont {Kubo}, \citenamefont
  {Kawano},\ and\ \citenamefont {Yamauchi}}]{munakata1992}%
  \BibitemOpen
  \bibfield  {author} {\bibinfo {author} {\bibfnamefont {F.}~\bibnamefont
  {Munakata}}, \bibinfo {author} {\bibfnamefont {K.}~\bibnamefont {Matsuura}},
  \bibinfo {author} {\bibfnamefont {K.}~\bibnamefont {Kubo}}, \bibinfo {author}
  {\bibfnamefont {T.}~\bibnamefont {Kawano}},\ and\ \bibinfo {author}
  {\bibfnamefont {H.}~\bibnamefont {Yamauchi}},\ }\bibfield  {title} {\bibinfo
  {title} {Thermoelectric power of
  {Bi}$_{2}${Sr}$_{2}${Ca}$_{1-x}${Y}$_{x}${Cu}$_{2}${O}$_{8+y}$},\ }\href
  {https://link.aps.org/doi/10.1103/PhysRevB.45.10604} {\bibfield  {journal}
  {\bibinfo  {journal} {Physical Review B}\ }\textbf {\bibinfo {volume} {45}},\
  \bibinfo {pages} {10604} (\bibinfo {year} {1992})}\BibitemShut {NoStop}%
\bibitem [{\citenamefont {Elizarova}\ and\ \citenamefont
  {Gasumyants}(2000)}]{elizarova2000}%
  \BibitemOpen
  \bibfield  {author} {\bibinfo {author} {\bibfnamefont {M.~V.}\ \bibnamefont
  {Elizarova}}\ and\ \bibinfo {author} {\bibfnamefont {V.~E.}\ \bibnamefont
  {Gasumyants}},\ }\bibfield  {title} {\bibinfo {title} {Band spectrum
  transformation and ${T}_c$ variation in the {La}$_{2-x}${Sr}$_x${CuO}$_y$
  system in the underdoped and overdoped regimes},\ }\href
  {https://journals.aps.org/prb/abstract/10.1103/PhysRevB.62.5989} {\bibfield
  {journal} {\bibinfo  {journal} {Physical Review B}\ }\textbf {\bibinfo
  {volume} {62}},\ \bibinfo {pages} {5989} (\bibinfo {year}
  {2000})}\BibitemShut {NoStop}%
\bibitem [{\citenamefont {Yamamoto}\ \emph {et~al.}(2000)\citenamefont
  {Yamamoto}, \citenamefont {Hu},\ and\ \citenamefont {Tajima}}]{yamamoto2000}%
  \BibitemOpen
  \bibfield  {author} {\bibinfo {author} {\bibfnamefont {A.}~\bibnamefont
  {Yamamoto}}, \bibinfo {author} {\bibfnamefont {W.-Z.}\ \bibnamefont {Hu}},\
  and\ \bibinfo {author} {\bibfnamefont {S.}~\bibnamefont {Tajima}},\
  }\bibfield  {title} {\bibinfo {title} {Thermoelectric power and resistivity
  of {Hg}{Ba}$_2${CuO}$_{4+\delta}$ over a wide doping range},\ }\href
  {https://doi.org/10.1103/PhysRevB.63.024504} {\bibfield  {journal} {\bibinfo
  {journal} {Physical Review B}\ }\textbf {\bibinfo {volume} {63}},\ \bibinfo
  {pages} {024504} (\bibinfo {year} {2000})}\BibitemShut {NoStop}%
\bibitem [{\citenamefont {Horio}\ \emph {et~al.}(2018)\citenamefont {Horio},
  \citenamefont {Hauser}, \citenamefont {Sassa}, \citenamefont {Mingazheva},
  \citenamefont {Sutter}, \citenamefont {Kramer}, \citenamefont {Cook},
  \citenamefont {Nocerino}, \citenamefont {Forslund}, \citenamefont
  {Tjernberg}, \citenamefont {Kobayashi}, \citenamefont {Chikina},
  \citenamefont {Schr{\"o}ter}, \citenamefont {Krieger}, \citenamefont
  {Schmitt}, \citenamefont {Strocov}, \citenamefont {Pyon}, \citenamefont
  {Takayama}, \citenamefont {Takagi}, \citenamefont {Lipscombe}, \citenamefont
  {Hayden}, \citenamefont {Ishikado}, \citenamefont {Eisaki}, \citenamefont
  {Neupert}, \citenamefont {M{\aa}nsson}, \citenamefont {Matt},\ and\
  \citenamefont {Chang}}]{horio2018}%
  \BibitemOpen
  \bibfield  {author} {\bibinfo {author} {\bibfnamefont {M.}~\bibnamefont
  {Horio}}, \bibinfo {author} {\bibfnamefont {K.}~\bibnamefont {Hauser}},
  \bibinfo {author} {\bibfnamefont {Y.}~\bibnamefont {Sassa}}, \bibinfo
  {author} {\bibfnamefont {Z.}~\bibnamefont {Mingazheva}}, \bibinfo {author}
  {\bibfnamefont {D.}~\bibnamefont {Sutter}}, \bibinfo {author} {\bibfnamefont
  {K.}~\bibnamefont {Kramer}}, \bibinfo {author} {\bibfnamefont
  {A.}~\bibnamefont {Cook}}, \bibinfo {author} {\bibfnamefont {E.}~\bibnamefont
  {Nocerino}}, \bibinfo {author} {\bibfnamefont {O.}~\bibnamefont {Forslund}},
  \bibinfo {author} {\bibfnamefont {O.}~\bibnamefont {Tjernberg}}, \bibinfo
  {author} {\bibfnamefont {M.}~\bibnamefont {Kobayashi}}, \bibinfo {author}
  {\bibfnamefont {A.}~\bibnamefont {Chikina}}, \bibinfo {author} {\bibfnamefont
  {N.}~\bibnamefont {Schr{\"o}ter}}, \bibinfo {author} {\bibfnamefont
  {J.}~\bibnamefont {Krieger}}, \bibinfo {author} {\bibfnamefont
  {T.}~\bibnamefont {Schmitt}}, \bibinfo {author} {\bibfnamefont
  {V.}~\bibnamefont {Strocov}}, \bibinfo {author} {\bibfnamefont
  {S.}~\bibnamefont {Pyon}}, \bibinfo {author} {\bibfnamefont {T.}~\bibnamefont
  {Takayama}}, \bibinfo {author} {\bibfnamefont {H.}~\bibnamefont {Takagi}},
  \bibinfo {author} {\bibfnamefont {O.}~\bibnamefont {Lipscombe}}, \bibinfo
  {author} {\bibfnamefont {S.}~\bibnamefont {Hayden}}, \bibinfo {author}
  {\bibfnamefont {M.}~\bibnamefont {Ishikado}}, \bibinfo {author}
  {\bibfnamefont {H.}~\bibnamefont {Eisaki}}, \bibinfo {author} {\bibfnamefont
  {T.}~\bibnamefont {Neupert}}, \bibinfo {author} {\bibfnamefont
  {M.}~\bibnamefont {M{\aa}nsson}}, \bibinfo {author} {\bibfnamefont
  {C.}~\bibnamefont {Matt}},\ and\ \bibinfo {author} {\bibfnamefont
  {J.}~\bibnamefont {Chang}},\ }\bibfield  {title} {\bibinfo {title}
  {Three-{Dimensional} {Fermi} {Surface} of {Overdoped} {La}-{Based}
  {Cuprates}},\ }\href {https://doi.org/10.1103/PhysRevLett.121.077004}
  {\bibfield  {journal} {\bibinfo  {journal} {Physical Review Letters}\
  }\textbf {\bibinfo {volume} {121}},\ \bibinfo {pages} {077004} (\bibinfo
  {year} {2018})}\BibitemShut {NoStop}%
\bibitem [{\citenamefont {Varma}\ \emph {et~al.}(1989)\citenamefont {Varma},
  \citenamefont {Littlewood}, \citenamefont {Schmitt-Rink}, \citenamefont
  {Abrahams},\ and\ \citenamefont {Ruckenstein}}]{varma1989}%
  \BibitemOpen
  \bibfield  {author} {\bibinfo {author} {\bibfnamefont {C.~M.}\ \bibnamefont
  {Varma}}, \bibinfo {author} {\bibfnamefont {P.~B.}\ \bibnamefont
  {Littlewood}}, \bibinfo {author} {\bibfnamefont {S.}~\bibnamefont
  {Schmitt-Rink}}, \bibinfo {author} {\bibfnamefont {E.}~\bibnamefont
  {Abrahams}},\ and\ \bibinfo {author} {\bibfnamefont {A.~E.}\ \bibnamefont
  {Ruckenstein}},\ }\bibfield  {title} {\bibinfo {title} {Phenomenology of the
  normal state of {Cu}-{O} high-temperature superconductors},\ }\href
  {https://doi.org/10.1103/PhysRevLett.63.1996} {\bibfield  {journal} {\bibinfo
   {journal} {Physical Review Letters}\ }\textbf {\bibinfo {volume} {63}},\
  \bibinfo {pages} {1996} (\bibinfo {year} {1989})}\BibitemShut {NoStop}%
\bibitem [{\citenamefont {Varma}(2020)}]{varma2020}%
  \BibitemOpen
  \bibfield  {author} {\bibinfo {author} {\bibfnamefont {C.~M.}\ \bibnamefont
  {Varma}},\ }\bibfield  {title} {\bibinfo {title} {Colloquium: {Linear} in
  temperature resistivity and associated mysteries including high temperature
  superconductivity},\ }\href {https://doi.org/10.1103/RevModPhys.92.031001}
  {\bibfield  {journal} {\bibinfo  {journal} {Reviews of Modern Physics}\
  }\textbf {\bibinfo {volume} {92}},\ \bibinfo {pages} {031001} (\bibinfo
  {year} {2020})}\BibitemShut {NoStop}%
\bibitem [{\citenamefont {Legros}\ \emph {et~al.}(2019)\citenamefont {Legros},
  \citenamefont {Benhabib}, \citenamefont {Tabis}, \citenamefont
  {Lalibert{\'e}}, \citenamefont {Dion}, \citenamefont {Lizaire}, \citenamefont
  {Vignolle}, \citenamefont {Vignolles}, \citenamefont {Raffy}, \citenamefont
  {Li}, \citenamefont {Auban-Senzier}, \citenamefont {Doiron-Leyraud},
  \citenamefont {Fournier}, \citenamefont {Colson}, \citenamefont {Taillefer},\
  and\ \citenamefont {Proust}}]{legros2019}%
  \BibitemOpen
  \bibfield  {author} {\bibinfo {author} {\bibfnamefont {A.}~\bibnamefont
  {Legros}}, \bibinfo {author} {\bibfnamefont {S.}~\bibnamefont {Benhabib}},
  \bibinfo {author} {\bibfnamefont {W.}~\bibnamefont {Tabis}}, \bibinfo
  {author} {\bibfnamefont {F.}~\bibnamefont {Lalibert{\'e}}}, \bibinfo {author}
  {\bibfnamefont {M.}~\bibnamefont {Dion}}, \bibinfo {author} {\bibfnamefont
  {M.}~\bibnamefont {Lizaire}}, \bibinfo {author} {\bibfnamefont
  {B.}~\bibnamefont {Vignolle}}, \bibinfo {author} {\bibfnamefont
  {D.}~\bibnamefont {Vignolles}}, \bibinfo {author} {\bibfnamefont
  {H.}~\bibnamefont {Raffy}}, \bibinfo {author} {\bibfnamefont {Z.~Z.}\
  \bibnamefont {Li}}, \bibinfo {author} {\bibfnamefont {P.}~\bibnamefont
  {Auban-Senzier}}, \bibinfo {author} {\bibfnamefont {N.}~\bibnamefont
  {Doiron-Leyraud}}, \bibinfo {author} {\bibfnamefont {P.}~\bibnamefont
  {Fournier}}, \bibinfo {author} {\bibfnamefont {D.}~\bibnamefont {Colson}},
  \bibinfo {author} {\bibfnamefont {L.}~\bibnamefont {Taillefer}},\ and\
  \bibinfo {author} {\bibfnamefont {C.}~\bibnamefont {Proust}},\ }\bibfield
  {title} {\bibinfo {title} {Universal ${T}$-linear resistivity and {Planckian}
  dissipation in overdoped cuprates},\ }\href
  {https://doi.org/10.1038/s41567-018-0334-2} {\bibfield  {journal} {\bibinfo
  {journal} {Nature Physics}\ }\textbf {\bibinfo {volume} {15}},\ \bibinfo
  {pages} {142} (\bibinfo {year} {2019})}\BibitemShut {NoStop}%
\bibitem [{\citenamefont {Bruin}\ \emph {et~al.}(2013)\citenamefont {Bruin},
  \citenamefont {Sakai}, \citenamefont {Perry},\ and\ \citenamefont
  {Mackenzie}}]{bruin2013}%
  \BibitemOpen
  \bibfield  {author} {\bibinfo {author} {\bibfnamefont {J.~A.~N.}\
  \bibnamefont {Bruin}}, \bibinfo {author} {\bibfnamefont {H.}~\bibnamefont
  {Sakai}}, \bibinfo {author} {\bibfnamefont {R.~S.}\ \bibnamefont {Perry}},\
  and\ \bibinfo {author} {\bibfnamefont {A.~P.}\ \bibnamefont {Mackenzie}},\
  }\bibfield  {title} {\bibinfo {title} {Similarity of scattering rates in
  metals showing ${T}$-linear resistivity},\ }\href
  {https://science.sciencemag.org/content/339/6121/804} {\bibfield  {journal}
  {\bibinfo  {journal} {Science}\ }\textbf {\bibinfo {volume} {339}},\ \bibinfo
  {pages} {804} (\bibinfo {year} {2013})}\BibitemShut {NoStop}%
\bibitem [{\citenamefont {Georges}\ and\ \citenamefont
  {Mravlje}(2021)}]{georges2021}%
  \BibitemOpen
  \bibfield  {author} {\bibinfo {author} {\bibfnamefont {A.}~\bibnamefont
  {Georges}}\ and\ \bibinfo {author} {\bibfnamefont {J.}~\bibnamefont
  {Mravlje}},\ }\bibfield  {title} {\bibinfo {title} {Skewed non-{Fermi}
  liquids and the {Seebeck} effect},\ }\href
  {https://doi.org/10.1103/PhysRevResearch.3.043132} {\bibfield  {journal}
  {\bibinfo  {journal} {Physical Review Research}\ }\textbf {\bibinfo {volume}
  {3}},\ \bibinfo {pages} {043132} (\bibinfo {year} {2021})}\BibitemShut
  {NoStop}%
\bibitem [{\citenamefont {Haule}\ and\ \citenamefont
  {Kotliar}(2009)}]{haule_2009}%
  \BibitemOpen
  \bibfield  {author} {\bibinfo {author} {\bibfnamefont {K.}~\bibnamefont
  {Haule}}\ and\ \bibinfo {author} {\bibfnamefont {G.}~\bibnamefont
  {Kotliar}},\ }\bibfield  {title} {\bibinfo {title} {Thermoelectrics near the
  {Mott} localization—delocalization transition},\ }\href
  {https://doi.org/10.1007/978-90-481-2892-1_7} {\bibfield  {journal} {\bibinfo
   {journal} {Properties and Applications of Thermoelectric Materials}\ ,\
  \bibinfo {pages} {119–131}} (\bibinfo {year} {2009})},\ \Eprint
  {https://arxiv.org/abs/arXiv:0907.0192} {arXiv:0907.0192} \BibitemShut
  {NoStop}%
\bibitem [{\citenamefont {Paul}\ and\ \citenamefont
  {Kotliar}(2001)}]{paul_kotliar_2001}%
  \BibitemOpen
  \bibfield  {author} {\bibinfo {author} {\bibfnamefont {I.}~\bibnamefont
  {Paul}}\ and\ \bibinfo {author} {\bibfnamefont {G.}~\bibnamefont {Kotliar}},\
  }\bibfield  {title} {\bibinfo {title} {Thermoelectric behavior near the
  magnetic quantum critical point},\ }\href
  {https://doi.org/10.1103/PhysRevB.64.184414} {\bibfield  {journal} {\bibinfo
  {journal} {Phys. Rev. B}\ }\textbf {\bibinfo {volume} {64}},\ \bibinfo
  {pages} {184414} (\bibinfo {year} {2001})}\BibitemShut {NoStop}%
\bibitem [{\citenamefont {Jin}\ \emph {et~al.}(2021)\citenamefont {Jin},
  \citenamefont {Narduzzo}, \citenamefont {Nohara}, \citenamefont {Takagi},
  \citenamefont {Hussey},\ and\ \citenamefont {Behnia}}]{jin2021}%
  \BibitemOpen
  \bibfield  {author} {\bibinfo {author} {\bibfnamefont {H.}~\bibnamefont
  {Jin}}, \bibinfo {author} {\bibfnamefont {A.}~\bibnamefont {Narduzzo}},
  \bibinfo {author} {\bibfnamefont {M.}~\bibnamefont {Nohara}}, \bibinfo
  {author} {\bibfnamefont {H.}~\bibnamefont {Takagi}}, \bibinfo {author}
  {\bibfnamefont {N.~E.}\ \bibnamefont {Hussey}},\ and\ \bibinfo {author}
  {\bibfnamefont {K.}~\bibnamefont {Behnia}},\ }\bibfield  {title} {\bibinfo
  {title} {Positive {Seebeck} coefficient in highly doped
  {La}$_{2-x}${Sr}$_x${CuO}$_4$ ($x$=0.33); its origin and implication},\
  }\href {https://doi.org/10.7566/JPSJ.90.053702} {\bibfield  {journal}
  {\bibinfo  {journal} {Journal of the Physical Society of Japan}\ }\textbf
  {\bibinfo {volume} {90}},\ \bibinfo {pages} {053702} (\bibinfo {year}
  {2021})}\BibitemShut {NoStop}%
\bibitem [{\citenamefont {Chang}\ \emph {et~al.}(2013)\citenamefont {Chang},
  \citenamefont {Månsson}, \citenamefont {Pailhès}, \citenamefont {Claesson},
  \citenamefont {Lipscombe}, \citenamefont {Hayden}, \citenamefont {Patthey},
  \citenamefont {Tjernberg},\ and\ \citenamefont {Mesot}}]{chang2013}%
  \BibitemOpen
  \bibfield  {author} {\bibinfo {author} {\bibfnamefont {J.}~\bibnamefont
  {Chang}}, \bibinfo {author} {\bibfnamefont {M.}~\bibnamefont {Månsson}},
  \bibinfo {author} {\bibfnamefont {S.}~\bibnamefont {Pailhès}}, \bibinfo
  {author} {\bibfnamefont {T.}~\bibnamefont {Claesson}}, \bibinfo {author}
  {\bibfnamefont {O.~J.}\ \bibnamefont {Lipscombe}}, \bibinfo {author}
  {\bibfnamefont {S.~M.}\ \bibnamefont {Hayden}}, \bibinfo {author}
  {\bibfnamefont {L.}~\bibnamefont {Patthey}}, \bibinfo {author} {\bibfnamefont
  {O.}~\bibnamefont {Tjernberg}},\ and\ \bibinfo {author} {\bibfnamefont
  {J.}~\bibnamefont {Mesot}},\ }\bibfield  {title} {\bibinfo {title}
  {Anisotropic breakdown of {Fermi} liquid quasiparticle excitations in
  overdoped {La}$_{2-x}${Sr}$_x${CuO}$_4$},\ }\href
  {https://doi.org/10.1038/ncomms3559} {\bibfield  {journal} {\bibinfo
  {journal} {Nature Communications}\ }\textbf {\bibinfo {volume} {4}},\
  \bibinfo {pages} {2559} (\bibinfo {year} {2013})}\BibitemShut {NoStop}%
\bibitem [{\citenamefont {Storey}\ \emph {et~al.}(2013)\citenamefont {Storey},
  \citenamefont {Tallon},\ and\ \citenamefont {Williams}}]{storey2013}%
  \BibitemOpen
  \bibfield  {author} {\bibinfo {author} {\bibfnamefont {J.~G.}\ \bibnamefont
  {Storey}}, \bibinfo {author} {\bibfnamefont {J.~L.}\ \bibnamefont {Tallon}},\
  and\ \bibinfo {author} {\bibfnamefont {G.~V.~M.}\ \bibnamefont {Williams}},\
  }\bibfield  {title} {\bibinfo {title} {Electron pockets and pseudogap
  asymmetry observed in the thermopower of underdoped cuprates},\ }\href
  {https://doi.org/10.1209/0295-5075/102/37006} {\bibfield  {journal} {\bibinfo
   {journal} {EPL}\ }\textbf {\bibinfo {volume} {102}},\ \bibinfo {pages}
  {37006} (\bibinfo {year} {2013})}\BibitemShut {NoStop}%
\bibitem [{\citenamefont {Abdel-Jawad}\ \emph {et~al.}(2006)\citenamefont
  {Abdel-Jawad}, \citenamefont {Kennett}, \citenamefont {Balicas},
  \citenamefont {Carrington}, \citenamefont {Mackenzie}, \citenamefont
  {McKenzie},\ and\ \citenamefont {Hussey}}]{abdeljawad2006}%
  \BibitemOpen
  \bibfield  {author} {\bibinfo {author} {\bibfnamefont {M.}~\bibnamefont
  {Abdel-Jawad}}, \bibinfo {author} {\bibfnamefont {M.~P.}\ \bibnamefont
  {Kennett}}, \bibinfo {author} {\bibfnamefont {L.}~\bibnamefont {Balicas}},
  \bibinfo {author} {\bibfnamefont {A.}~\bibnamefont {Carrington}}, \bibinfo
  {author} {\bibfnamefont {A.~P.}\ \bibnamefont {Mackenzie}}, \bibinfo {author}
  {\bibfnamefont {R.~H.}\ \bibnamefont {McKenzie}},\ and\ \bibinfo {author}
  {\bibfnamefont {N.~E.}\ \bibnamefont {Hussey}},\ }\bibfield  {title}
  {\bibinfo {title} {Anisotropic scattering and anomalous normal-state
  transport in a high-temperature superconductor},\ }\href
  {https://doi.org/10.1038/nphys449} {\bibfield  {journal} {\bibinfo  {journal}
  {Nature Physics}\ }\textbf {\bibinfo {volume} {2}},\ \bibinfo {pages} {821}
  (\bibinfo {year} {2006})}\BibitemShut {NoStop}%
\bibitem [{\citenamefont {Yoshida}\ \emph {et~al.}(2007)\citenamefont
  {Yoshida}, \citenamefont {Zhou}, \citenamefont {Lu}, \citenamefont {Komiya},
  \citenamefont {Ando}, \citenamefont {Eisaki}, \citenamefont {Kakeshita},
  \citenamefont {Uchida}, \citenamefont {Hussain}, \citenamefont {Shen},\ and\
  \citenamefont {Fujimori}}]{yoshida_2007}%
  \BibitemOpen
  \bibfield  {author} {\bibinfo {author} {\bibfnamefont {T.}~\bibnamefont
  {Yoshida}}, \bibinfo {author} {\bibfnamefont {X.~J.}\ \bibnamefont {Zhou}},
  \bibinfo {author} {\bibfnamefont {D.~H.}\ \bibnamefont {Lu}}, \bibinfo
  {author} {\bibfnamefont {S.}~\bibnamefont {Komiya}}, \bibinfo {author}
  {\bibfnamefont {Y.}~\bibnamefont {Ando}}, \bibinfo {author} {\bibfnamefont
  {H.}~\bibnamefont {Eisaki}}, \bibinfo {author} {\bibfnamefont
  {T.}~\bibnamefont {Kakeshita}}, \bibinfo {author} {\bibfnamefont
  {S.}~\bibnamefont {Uchida}}, \bibinfo {author} {\bibfnamefont
  {Z.}~\bibnamefont {Hussain}}, \bibinfo {author} {\bibfnamefont {Z.-X.}\
  \bibnamefont {Shen}},\ and\ \bibinfo {author} {\bibfnamefont
  {A.}~\bibnamefont {Fujimori}},\ }\bibfield  {title} {\bibinfo {title}
  {Low-energy electronic structure of the high-${T}_c$ cuprates
  {La}$_{2-x}${Sr}$_x${CuO}$_4$ studied by angle-resolved photoemission
  spectroscopy},\ }\href {https://doi.org/10.1088/0953-8984/19/12/125209}
  {\bibfield  {journal} {\bibinfo  {journal} {Journal of Physics: Condensed
  Matter}\ }\textbf {\bibinfo {volume} {19}},\ \bibinfo {pages} {125209}
  (\bibinfo {year} {2007})}\BibitemShut {NoStop}%
\bibitem [{\citenamefont {Shahbazi}\ and\ \citenamefont
  {Bourbonnais}(2016)}]{shahbazi2016}%
  \BibitemOpen
  \bibfield  {author} {\bibinfo {author} {\bibfnamefont {M.}~\bibnamefont
  {Shahbazi}}\ and\ \bibinfo {author} {\bibfnamefont {C.}~\bibnamefont
  {Bourbonnais}},\ }\bibfield  {title} {\bibinfo {title} {Seebeck coefficient
  in correlated low-dimensional organic metals},\ }\href
  {https://doi.org/10.1103/PhysRevB.94.195153} {\bibfield  {journal} {\bibinfo
  {journal} {Phys. Rev. B}\ }\textbf {\bibinfo {volume} {94}},\ \bibinfo
  {pages} {195153} (\bibinfo {year} {2016})}\BibitemShut {NoStop}%
\bibitem [{\citenamefont {Gourgout}\ \emph {et~al.}(2021)\citenamefont
  {Gourgout}, \citenamefont {Ataei}, \citenamefont {Boulanger}, \citenamefont
  {Badoux}, \citenamefont {Th\'eriault}, \citenamefont {Graf}, \citenamefont
  {Zhou}, \citenamefont {Pyon}, \citenamefont {Takayama}, \citenamefont
  {Takagi}, \citenamefont {Doiron-Leyraud},\ and\ \citenamefont
  {Taillefer}}]{gourgout2021pressure}%
  \BibitemOpen
  \bibfield  {author} {\bibinfo {author} {\bibfnamefont {A.}~\bibnamefont
  {Gourgout}}, \bibinfo {author} {\bibfnamefont {A.}~\bibnamefont {Ataei}},
  \bibinfo {author} {\bibfnamefont {M.-E.}\ \bibnamefont {Boulanger}}, \bibinfo
  {author} {\bibfnamefont {S.}~\bibnamefont {Badoux}}, \bibinfo {author}
  {\bibfnamefont {S.}~\bibnamefont {Th\'eriault}}, \bibinfo {author}
  {\bibfnamefont {D.}~\bibnamefont {Graf}}, \bibinfo {author} {\bibfnamefont
  {J.-S.}\ \bibnamefont {Zhou}}, \bibinfo {author} {\bibfnamefont
  {S.}~\bibnamefont {Pyon}}, \bibinfo {author} {\bibfnamefont {T.}~\bibnamefont
  {Takayama}}, \bibinfo {author} {\bibfnamefont {H.}~\bibnamefont {Takagi}},
  \bibinfo {author} {\bibfnamefont {N.}~\bibnamefont {Doiron-Leyraud}},\ and\
  \bibinfo {author} {\bibfnamefont {L.}~\bibnamefont {Taillefer}},\ }\bibfield
  {title} {\bibinfo {title} {Effect of pressure on the pseudogap and charge
  density wave phases of the cuprate {Nd}-{L}{S}{C}{O} probed by thermopower
  measurements},\ }\href {https://doi.org/10.1103/PhysRevResearch.3.023066}
  {\bibfield  {journal} {\bibinfo  {journal} {Phys. Rev. Research}\ }\textbf
  {\bibinfo {volume} {3}},\ \bibinfo {pages} {023066} (\bibinfo {year}
  {2021})}\BibitemShut {NoStop}%
\bibitem [{\citenamefont {Deng}\ \emph {et~al.}(2013)\citenamefont {Deng},
  \citenamefont {Mravlje}, \citenamefont {\ifmmode~\check{Z}\else
  \v{Z}\fi{}itko}, \citenamefont {Ferrero}, \citenamefont {Kotliar},\ and\
  \citenamefont {Georges}}]{deng_2013}%
  \BibitemOpen
  \bibfield  {author} {\bibinfo {author} {\bibfnamefont {X.}~\bibnamefont
  {Deng}}, \bibinfo {author} {\bibfnamefont {J.}~\bibnamefont {Mravlje}},
  \bibinfo {author} {\bibfnamefont {R.}~\bibnamefont {\ifmmode~\check{Z}\else
  \v{Z}\fi{}itko}}, \bibinfo {author} {\bibfnamefont {M.}~\bibnamefont
  {Ferrero}}, \bibinfo {author} {\bibfnamefont {G.}~\bibnamefont {Kotliar}},\
  and\ \bibinfo {author} {\bibfnamefont {A.}~\bibnamefont {Georges}},\
  }\bibfield  {title} {\bibinfo {title} {How bad metals turn good:
  Spectroscopic signatures of resilient quasiparticles},\ }\href
  {https://doi.org/10.1103/PhysRevLett.110.086401} {\bibfield  {journal}
  {\bibinfo  {journal} {Phys. Rev. Lett.}\ }\textbf {\bibinfo {volume} {110}},\
  \bibinfo {pages} {086401} (\bibinfo {year} {2013})}\BibitemShut {NoStop}%
\bibitem [{\citenamefont {Xu}\ \emph {et~al.}(2013)\citenamefont {Xu},
  \citenamefont {Haule},\ and\ \citenamefont {Kotliar}}]{xu_haule_2013}%
  \BibitemOpen
  \bibfield  {author} {\bibinfo {author} {\bibfnamefont {W.}~\bibnamefont
  {Xu}}, \bibinfo {author} {\bibfnamefont {K.}~\bibnamefont {Haule}},\ and\
  \bibinfo {author} {\bibfnamefont {G.}~\bibnamefont {Kotliar}},\ }\bibfield
  {title} {\bibinfo {title} {Hidden fermi liquid, scattering rate saturation,
  and nernst effect: A dynamical mean-field theory perspective},\ }\href
  {https://doi.org/10.1103/PhysRevLett.111.036401} {\bibfield  {journal}
  {\bibinfo  {journal} {Phys. Rev. Lett.}\ }\textbf {\bibinfo {volume} {111}},\
  \bibinfo {pages} {036401} (\bibinfo {year} {2013})}\BibitemShut {NoStop}%
\end{thebibliography}
\end{document}